\documentclass{article}

\usepackage{arxiv}

\usepackage[utf8]{inputenc} 
\usepackage[T1]{fontenc}    
\usepackage{hyperref}       
\usepackage{url}            
\usepackage{booktabs}       
\usepackage{amsfonts}       
\usepackage{amsmath}        
\usepackage{nicefrac}       
\usepackage{microtype}      
\usepackage{placeins}       
\usepackage{placeins}       
\usepackage{natbib}         
\usepackage{lipsum}
\usepackage{graphicx}
\graphicspath{ {./images/} }

\title{Inferring processes within dynamic forest models using hybrid modeling}

\author{
 Maximilian Pichler$^*$ \\
 Theoretical Ecology \\
 University of Regensburg \\
 93053 Regensburg\\
 Germany \\
 \texttt{maximilian.pichler@ur.de} \\
 \And
 Yannek Käber$^*$ \\
 Biometry and Environmental System Analysis\\
 University of Freiburg\\
 79098 Freiburg\\
 Germany \\
 \texttt{yannek.kaeber@biom.uni-freiburg.de} \\
}

\begin{document}
\maketitle

\def\thefootnote{*}\footnotetext{These authors contributed equally to this work}\def\thefootnote{\arabic{footnote}}
\begin{abstract}
Modeling forest dynamics under novel climatic conditions requires a careful balance between process-based understanding and empirical flexibility. Dynamic Vegetation Models (DVM) represent ecological processes mechanistically, but their performance is prone to misspecified assumptions about functional forms. Inferring the structure of these processes and their functional forms correctly from data remains a major challenge because current approaches, such as plug-in estimators, have proven ineffective. We introduce Forest Informed Neural Networks (FINN), a hybrid modeling approach that combines a forest gap model with deep neural networks (DNN). FINN replaces processes with DNNs, which are then calibrated alongside the other mechanistic components in one unified step. In a case study on the Barro Colorado Island 50-ha plot we demonstrate that replacing the growth process with a DNN improves predictive performance and succession trajectories compared to a mechanistic version of FINN. Furthermore, we discovered that the DNN learned an ecologically plausible, improved functional form of the growth process, which we extracted from the DNN using explainable AI. In conclusion, our new hybrid modeling approach offers a versatile opportunity to infer forest dynamics from data and to improve forecasts of ecosystem trajectories under unprecedented environmental change.
\end{abstract}


\section*{Main}
Forest ecosystems play a central role in Earth’s carbon and water cycle, and in climate regulation, acting as key components in both local and global ecological systems. Understanding and accurately predicting how forests respond to environmental changes is crucial for addressing climate change, conservation, and resource management \citep{forzieriEmergingSignalsDeclining2022, mcdowellPervasiveShiftsForest2020}. To this end, we rely on Dynamic Vegetation Models (DVM) that serve as \textit{repositories of integrated ecosystem knowledge} \citep{botkinEcologicalConsequencesComputer1972} and simulate mechanistic processes, from small scale such as photosynthesis to large scale such as tree community composition across space and time \citep{bugmannEvolutionComplexityDiversity2022,moorcroftMethodScalingVegetation2001, seidlIndividualbasedProcessModel2012, smithRepresentationVegetationDynamics2001}.

However, the reliability of DVMs depends strongly on how well they represent the underlying processes \citep{bevenManifestoEquifinalityThesis2006, cailleret_bayesian_2020, kaberInferringTreeRegeneration2024, vanoijenBayesianCalibrationProcessbased2005}. Misspecified processes cause structural model bias, which decreases the accuracy of the DVM, especially when predicting outside the observed data \citep{oberprillerRobustStatisticalInference2021}. But choosing the correct functional form of a process has proven to be non-trivial. Existing functional forms are often educated guesses based on isolated empirical correlations that do not consider the broader context (i.e. other processes). Inverse calibration methods (manual or systematic) \citep{hartigStatisticalInferenceStochastic2011,hartigConnectingDynamicVegetation2012} are usually limited by the practical infeasibility of evaluating a large number of candidate models \citep[e.g.][]{delpierreChillingForcingTemperatures2019a, kaberInferringTreeRegeneration2024}. Even worse, structural bias often goes undetected because other processes can compensate for it, thereby obscuring its source and recovering predictive accuracy (i.e., predicting the right things for the wrong reasons) \citep{cameronIssuesCalibratingModels2022,oberprillerRobustStatisticalInference2021}. Therefore, an approach that can automatically infer the functional form of a process empirically from data would be much more appealing. However, this requires a highly flexible empirical model, as well as joint calibration with the other mechanistic components of a DVM.

To address this challenge, the empirical model of the process must be sufficiently flexible to approximate any realistic functional form. Recent advances in deep learning (DL), particularly deep neural networks (DNN), offer a promising solution \citep{pichlerMachineLearningDeep2023}. DNNs can approximate any nonlinear function and automatically capture complex interactions between variables \citep[e.g.][]{pichlerMachineLearningAlgorithms2020}, offering advantages over more rigid alternatives and reducing the need for cumbersome model selection or manual formulation. However, their high flexibility comes at the cost of non-interpretability \citep{pichlerCanPredictiveModels2023}, which is at odds with our interest in understanding the functional form learned by the DNN. Tools of explainable Artificial Intelligence (xAI) were developed to address this problem \citep{ryoExplainableArtificialIntelligence2021}. Explainable AI tools analyze how model predictions change with controlled permutations of the inputs, analogous to sensitivity analysis \citep{murdochXai, malchowCalibrationSensitivityUncertainty2024}. Together, the flexibility of DNNs and the post-hoc interpretability of xAI create a synergy that can advance scientific discovery \citep[e.g.][]{daviesAdvancingMathematicsGuiding2021}, which could include the derivation of forest processes.

While flexibility addresses the functional form of a misspecified process, it has been shown that joint calibration of all components (mechanistic or not) improves the reliability of DVM. Calibrating processes externally, without the same context (i.e., different inputs and not in the presence of the other processes), can yield misleading inferences, similar to spurious correlations in statistics \citep{woodPartiallySpecifiedEcological2001}. Even when sharing the same input-output structure, plug-in models of processes often perform worse than target processes calibrated jointly with other DVM processes \citep[e.g.][]{hulsmannHowKillTree2018, gonzalez_inverse_2016, cailleret_bayesian_2020}. This is likely because DVMs are designed to produce emergent patterns of forest dynamics \citep{moorcroftMethodScalingVegetation2001} where process-level observations are interrelated over time and with each other, which is difficult to replicate with external independent models. Another reason could be that, since our understanding of nature will never be perfect, structural model bias in DVM is probably unavoidable. This causes the processes to compensate for each other during joint calibration \citep{oberprillerRobustStatisticalInference2021}, resulting in better performance than external plug-in models of processes \citep[see also][]{cailleret_bayesian_2020}.

\begin{figure}
  \centering
  \includegraphics[width=.7\linewidth]{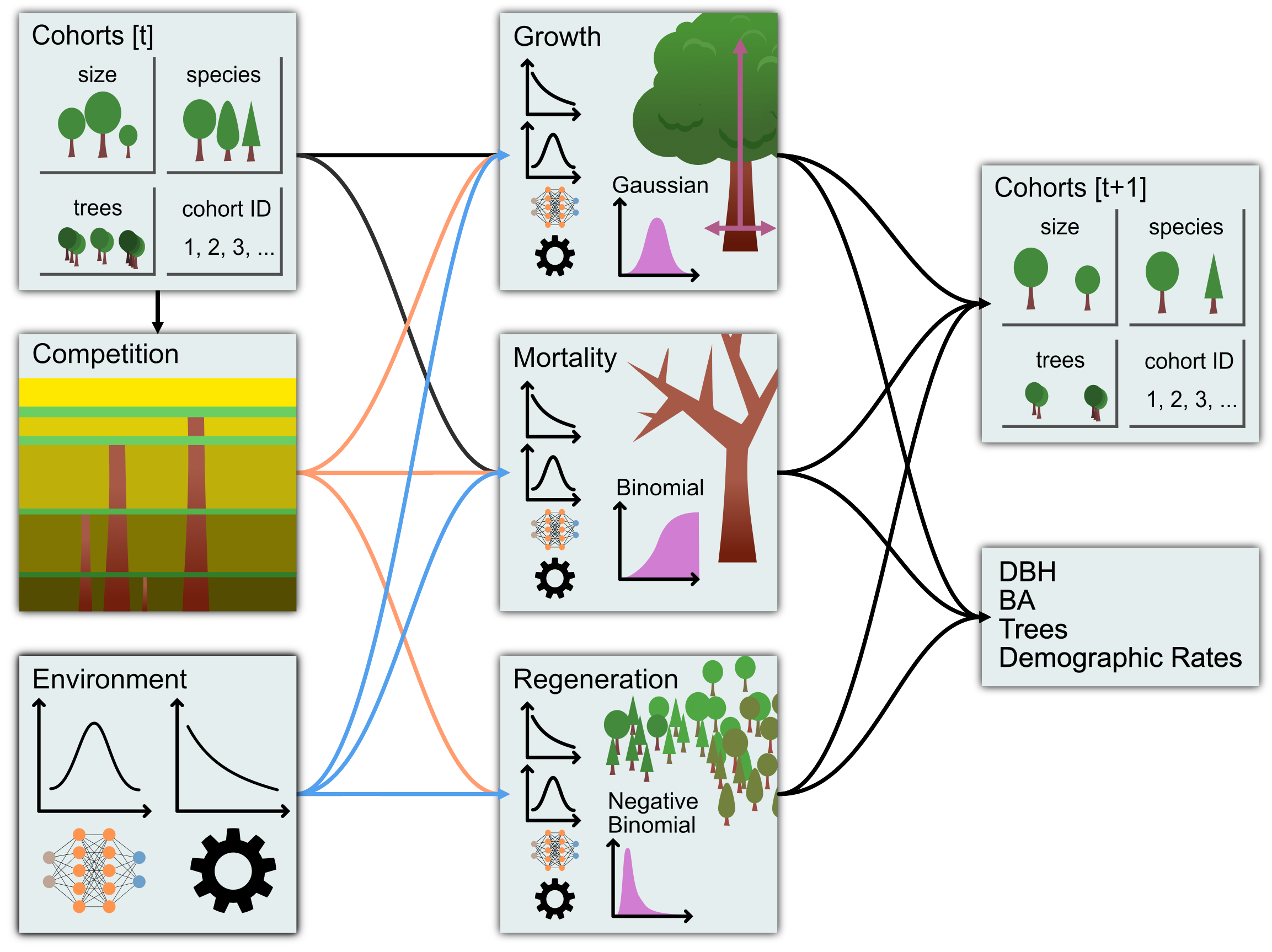}
  \caption{Overview of the Forest-Informed-Neural-Network (FINN) Model. FINN provides a single architecture that combines empirical data and process knowledge. FINN achieves this by replacing processes with a deep neural network (DNN) (Hybrid-FINN) as required. The DNN is jointly optimized with the rest of the DVM. Consequently, FINN only increases complexity in specific processes. FINN can be also inverse calibrated as a full process-based model (Process-FINN). At its core, FINN is a simple DVM which simulates forest dynamics (basal area (ba); diameter at breast height (dbh); number of trees (trees); demographic rates: growth rate, mortality rate, and recruitment) in discrete time steps (e.g. annually), where trees are grouped into cohorts characterized by size (e.g. dbh), number of trees, and species identity. The next forest state ($cohorts[t+1]$) is simulated from the initial forest conditions ($cohorts [t]$), taking into account competition and environmental factors that influence the demographic processes of growth, mortality and regeneration.}\label{fig:figure_1}
\end{figure}

To meet both criteria, we propose Forest Informed Neural Networks (FINN) (Fig.~\ref{fig:figure_1}), a DVM that replaces selected processes (e.g., growth, mortality, and regeneration) with DNNs (Fig.~\ref{fig:figure_1}). Unlike other hybrid modeling approaches with pre-calibrated plug-in models \citep{chenIterativeIntegrationDeep2023,reichsteinDeepLearningProcess2019}, FINN calibrates the replaced processes jointly with the remaining mechanistic processes end-to-end. This differs from process-informed neural networks, where process-based constraints adjust the model outputs \citep[][]{wesselkampProcessInformedNeuralNetworks2024}; in FINN the replaced processes are learned within and constrained by the surrounding DVM scaffold, which helps calibrate the DNN to predict emergent patterns of forest dynamics (Fig.~\ref{fig:figure_4}). Consequently, FINN can learn the functional form of processes from stand- and population-level without direct process-level observations. After calibration, processes approximated by a DNN can be explained by xAI (Fig.~\ref{fig:figure_5}). Consequently, FINN can learn the most suitable functional form for each process during a single, unified calibration step. This effectively combines classic inverse calibration and modern DL methods \citep[see also][]{boussangePartitioningTimeSeries2024,sapienzaDifferentiableProgrammingDifferential2024}.

The purpose of this study is to evaluate the potential of FINN. We tested FINN on simulated data as well as on the Barro Colorado Island (BCI) forest inventory data \citep{conditDemographicTrendsClimate2017} to assess the following questions. 
i) Can the FINN approach be calibrated to forest dynamic data and recover the true functional form? 
ii) What is the advantage of replacing a mechanistic process with a deep neural network? 
iii) Can we use the tools of explainable AI to understand the functional form learned by a DNN process within FINN?

\section*{The Forest-Informed Neural Network (FINN) model}

From the forest DVM perspective, FINN integrates a classical forest gap model with mechanistic and data-driven demographic process equations. Its core structure represents trees as cohorts within independent patches, where demographic processes – regeneration, growth, and mortality – operate alongside competition and environmental inputs to simulate forest dynamics in discrete timesteps. Each process in FINN depends on the environment independent of each other.

FINN can be a process model (Process-FINN) or a hybrid model (Hybrid-FINN) in which one or more processes are replaced by a DNN. Process-FINN is a mechanistic model with a predefined functional form of demographic processes (see methods for a detailed description of each process) and serves as a default. Hybrid-FINN refers to the same model but with at least one process being replaced by a DNN.

Simulations take place on independent patches following the classic forest gap model approach of JABOWA \citep{botkinEcologicalConsequencesComputer1972}. At each timestep $t$, cohorts $C_{t+1}$ are updated by demographic processes that depend on the environment $env_t$ and the current cohorts $C_t$:
\begin{align*}
C_{t+1} = f(\text{regeneration}(env_t, regPar_s, light_t), \\ \text{growth}(env_t, C_t, growthPar_s, light_t), \\ \text{mortality}(env_t, C_t, mortPar_s, light_t)) 
\end{align*}
where $regPar_s$, $growthPar_s$, and $mortPar_s$ are species-specific parameters. Each process also incorporates a measure of competitive pressure (light) to approximate competition for light, space, and other resources, which is computed via a competition function:
\begin{align*}
light_t = \text{competition}(C_t, compPar_s)
\end{align*}
where $C_t$ represents the cohorts in one patch at time $t$ and $compPar_s$ includes parameters for height allometry (\texttt{"height"}~\texttt{"dbh"}) and a scaling factor for the competitive strength of each species (i.e., the ability of a species to cast shade or suppress neighboring trees) (cf. Fig.~S1). A detailed model description is given in the Appendix.

In contrast to other DVMs, FINN is based on an architecture that is compatible with classical empirical calibration and DL methods for parameter estimation. Specifically, all inputs, states, outputs, and parameters are represented with arrays, and as a fully analytical, differentiable DVM. We ensured that all functions and processes are differentiable and avoided model structures that introduce insensitivity and non-differentiability in parameter space such as step functions. By that FINN allows gradient-based optimization for the inverse calibration of all its parameters via backpropagation \citep{rumelhartLearningRepresentationsBackpropagating1986} as gradient-based optimization can handle many parameters efficiently \citep{sapienzaDifferentiableProgrammingDifferential2024}. 

To achieve that, we implemented FINN within the Torch for R framework \citep{falbel_torch_2025}, a state-of-the-art DL framework that supports automatic differentiation (AD). Also, the DL framework allows to run FINN on graphical processing units (GPUs), which substantially reduces the computation time (Fig.~S11).

\section*{Results}
Traditional dynamic vegetation models (DVMs) integrate multiple processes into a single framework based on expert-driven formulations of these processes. In contrast, our goal is to extract as much information as possible directly from the data. To this end, we compare two approaches: a) a classical approach that involves the inverse calibration of a process model (hereafter referred to as Process-FINN), and b) a hybrid model in which we replaced the growth process by a deep neural network (DNN) to explore the possibilities of learning complex processes with DL (hereafter referred to as Hybrid-FINN) (Fig.~\ref{fig:figure_1}). We focus on growth because it is represented with similar functional forms across DVMs (i.e., saturating growth based on the interactions between radiation and photosynthesis), enabling clean comparison to classical models.

Demographic processes in DVM usually result from derived factors based on the functional form of the variable relations embedded in the model. Common ways to combine these factors are (weighted) product derivation or choosing the minimum (cf. Liebig's law of the minimum). Within Process-FINN growth is the product of three factors: shade, size-dependent growth, and a linear term representing environmental effects.

\begin{align*}
g_{i,p,c}
  =\text{light}_{i,p,c}^{(\text{growth})}\,
   \cdot\text{size}_{i,p,c}^{(\text{growth})}\,
   \cdot\text{env}_{i,s}^{(\text{growth})}
\end{align*}
whereas within Hybrid-FINN growth simplifies to
\begin{align*}
g_{i,p,c} = f(\text{dbh}_{i,p,c}, \text{trees}_{i,p,c}, \text{light}_{i,p,c}, \\ \text{species}_{i,p,c}, \text{env}_{i^{(\text{growth})},s})
\end{align*}
where each input variable serves directly as input to a DNN without making any assumptions about the functional form of the relations between variables.

\subsection*{Recovery of a functional form in Hybrid-FINN}
Before calibrating Hybrid-FINN on real data where the functional form of the processes is unknown, we tested whether Hybrid-FINN  can recover the functional form of a process at all using a simulated scenario (cf. Fig.~\ref{fig:figure_1}). 

\begin{figure}
\centering
\includegraphics[width=.99\linewidth]{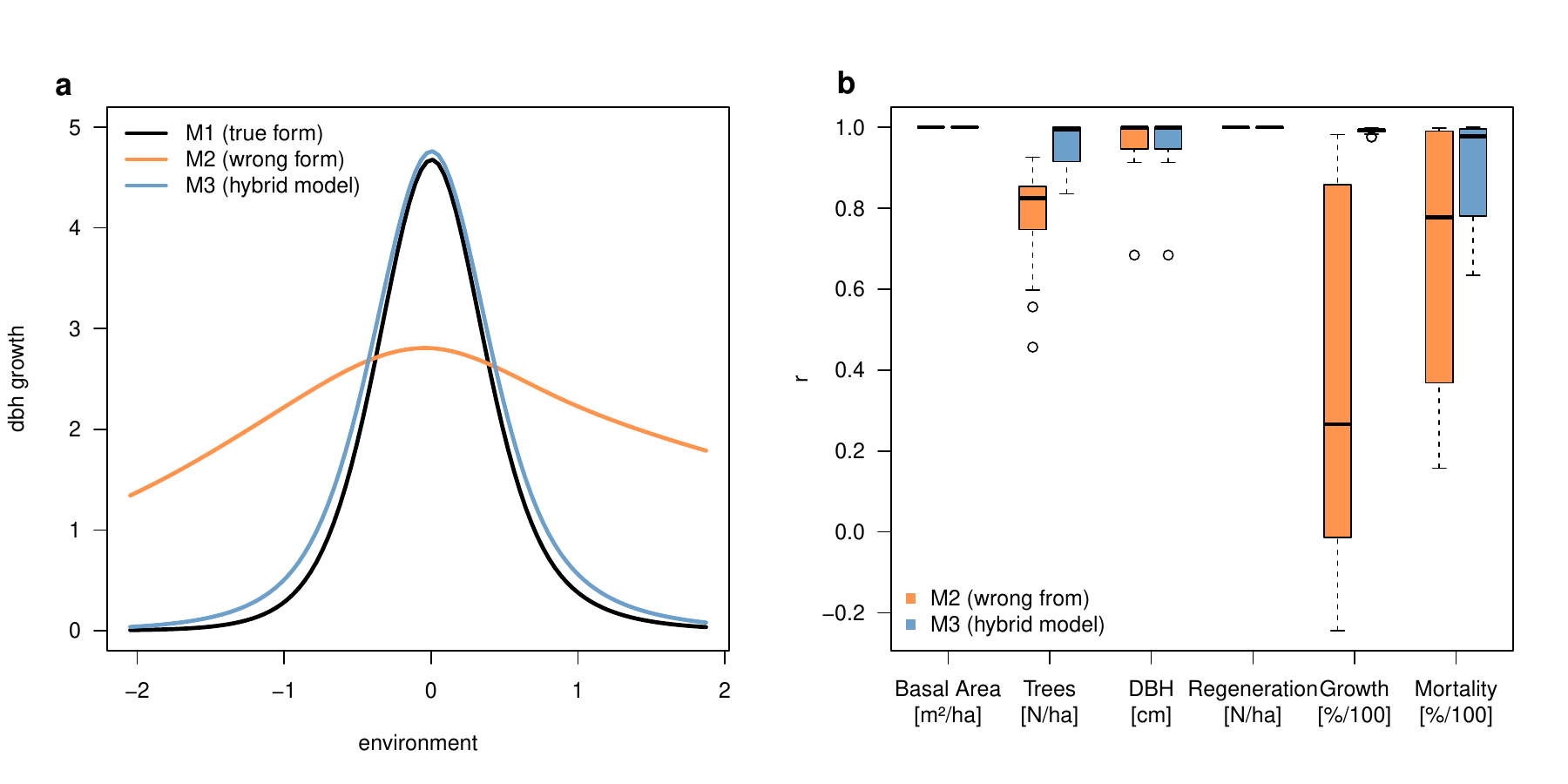}
\caption{Hybrid-FINN recovers the functional form of a process and improves predictive performance. To test whether Hybrid-FINN can recover the functional form of a process, we simulated data from a Process-FINN model (M1), in which the functional form of the growth process includes an environmental unimodal niche curve (quadratic effect of the environment). Then, we tested two models: M3, a Hybrid-FINN in which growth was replaced by a deep neural network; and M2, a misspecified Process-FINN model with an assumed linear relationship between environment and growth. (a) compares  the calibrated growth-environment relationships of models M2 and M3 to the true model M1. (b) compares the predictive performance of models M2 and M3 with the observed data, measured by the Spearman correlation factor. Correlations are based on equilibrium simulations and represent the relation for each variable and site over 500 simulated timesteps.}
\label{fig:figure_2}
\end{figure}

The true model was defined as a Process-FINN model for a single species with known parameters and a growth process that incorporates \textit{a quadratic response to the environment (true form, M1)} to simulate the dynamics of a single species (Fig.~S3). We calibrated two alternative FINN models to the simulated data: a Process-FINN model with a wrong growth formulation with a \textit{ linear response to environment (wrong form, M2)}, and a Hybrid-FINN model where a \textit{DNN replaces the growth process (hybrid model, M3)}.

We find that Hybrid-FINN (M3) successfully recovered the functional form of the true model (M1), while the misspecified model (M2) failed to capture the true form of the process (Fig.~\ref{fig:figure_2}a) and showed worse predictive performance of number of trees, growth rate, and mortality rate (Fig.~\ref{fig:figure_2}b).

\subsection*{Hybrid modeling of tree growth improves predictions and succession trajectories of forest dynamics}
Next, we tested whether hybrid modeling improves predictions of forest dynamics and of processes with real data. We compared the predictive performance of three inversely calibrated models, Hybrid-FINN in which a DNN replaced the growth process, Process-FINN, and a naïve Deep Neural Network (naïve DNN, 'naive' refers to missing explicit process knowledge) on the BCI data using five-fold spatially blocked cross-validation. The three models were inverse calibrated on the stand variables (ba, trees, dbh) and the demographic rates (growth rate, mortality rate, regeneration).

\begin{figure}
\centering
\includegraphics[width=.99\linewidth]{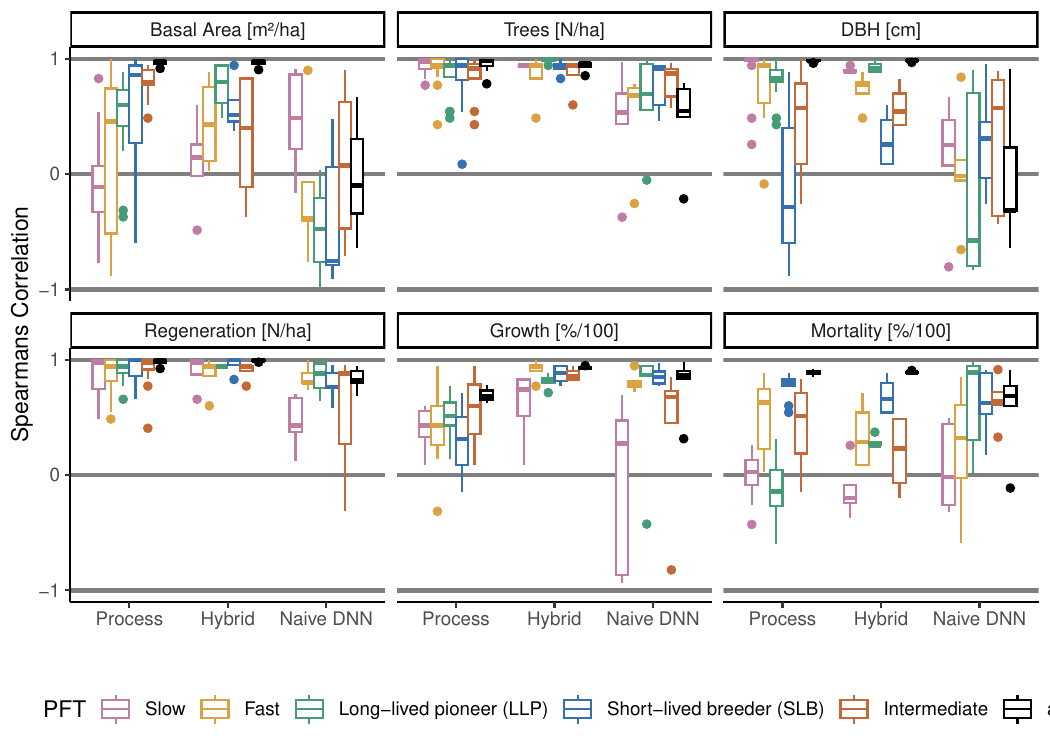}
\caption{Spearman correlation of observed and simulated variables for five PFTs from the five-fold blocked spatial cross-validation of all models calibrated/trained on the BCI data. 'Process' refers to Process-FINN as a full process model with fixed functional forms, 'Hybrid' refers to Hybrid-FINN in which the growth process is replaced by a DNN, 'Naïve NN' refers to a single DNN that was trained on the response variables for each time step. The naïve NN received the derived response variables of each previous time step as well as the current environment as input. PFTs are according to Rüger et al \citep{rugerDemographicTradeoffsPredict2020}.}
\label{fig:figure_3}
\end{figure}
Across all three modeling approaches, Hybrid-FINN delivered the best overall predictive performance (Fig.~\ref{fig:figure_3}). Process-FINN demonstrated small errors and high Spearman correlations above 0.9, while growth rate and mortality rate were less precise ($\rho \approx 0.4-0.8$) (Fig.~\ref{fig:figure_3}). The naïve neural network performed slightly better than Process-FINN for growth rate but worse for the stand variables ($\rho \approx 0.5-0.8$ for basal area and number of trees; $0.3–0.7$ for growth rate and mortality rate). By contrast, Hybrid-FINN showed higher Spearman correlations for basal area, dbh, number of trees and regeneration ($\rho > 0.9$) than the other two models. Most importantly, Hybrid-FINN achieved the highest Spearman correlation for growth rate, especially compared to Process-FINN ($\rho \approx 0.5$ to $\rho \approx 0.9$) (Fig.~\ref{fig:figure_3}). 

Another notable finding is that the predictions for individual PFTs were less accurate compared to all combined PFTs (Fig.~\ref{fig:figure_3}), indicating that the model captured the interspecific differences of species better than the variation within individual species, which might be explained by the lack of environmental variation in the data. 


\begin{figure*}[t!]
\centering
\includegraphics[width=.99\linewidth]{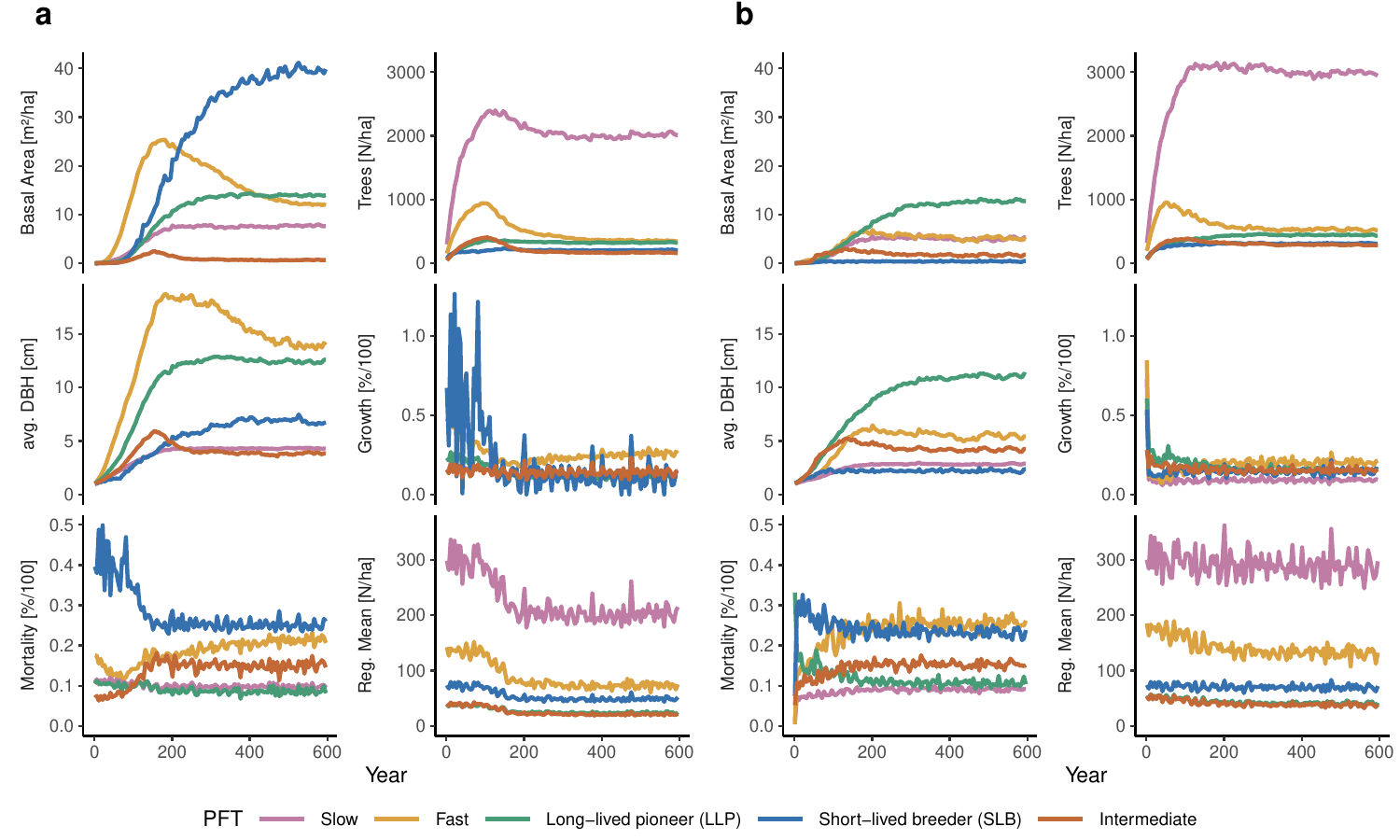}
\caption{Simulated successional trajectories of stand characteristics and demographic rates of five PFTs for Process-FINN (a) and Hybrid-FINN (b), where the growth process is replaced by a DNN. Note that rates are not annual and represent changes of five year intervals. Both models were trained with the BCI forest data with 7 censuses. Forest dynamics were simulated for 600 years. Disturbance regime of Rüger et al.\citep{rugerDemographicTradeoffsPredict2020} was used.} 
\label{fig:figure_4}
\end{figure*}

We find that the Hybrid-FINN model produces more plausible succession trajectories for the PFTs (cf. \citealp{rugerDemographicTradeoffsPredict2020}) than the Process-FINN model (run for 600 years, Fig.~\ref{fig:figure_4}). The resulting equilibrium of the Hybrid-FINN mirrors the observed dynamics of the BCI 50 ha plot \citep{conditDemographicTrendsClimate2017,condit_complete_2019,condit_census_2019} (same disturbance regime used by \citealp{rugerDemographicTradeoffsPredict2020}) (Fig.~\ref{fig:figure_4}b). However, Process-FINN failed to simulate reasonable patterns for the PFT ‘Short-lived breeder’ (SLB), resulting in an overall implausible equilibrium (Fig.~\ref{fig:figure_4}a).

\subsection*{Inferring the functional form between growth and light availability in Hybrid-FINN}
\begin{figure}
\centering
\includegraphics[width=.99\linewidth]{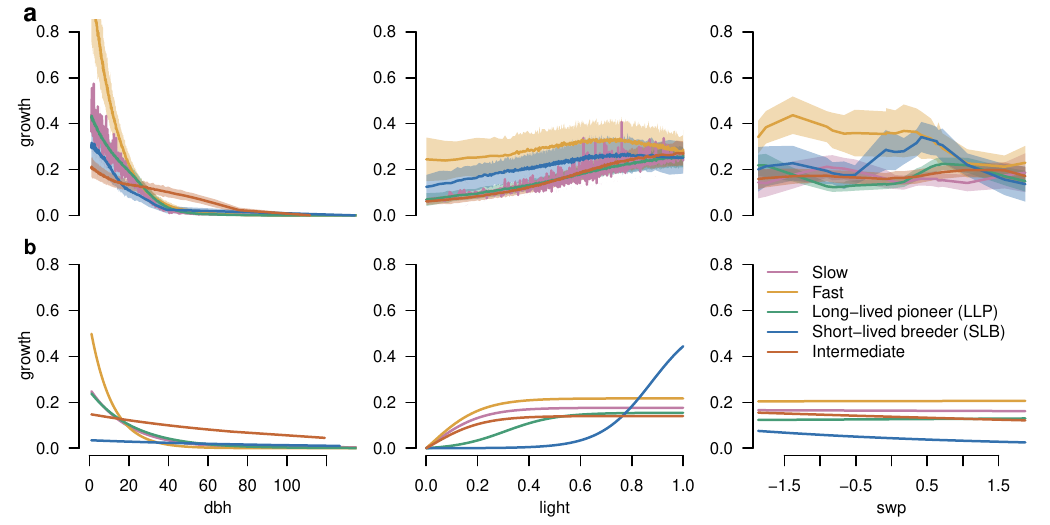}
\caption{Simulated relative growth response to available light and tree size (dbh) for 5 PFTs for the hybrid model (a) and the process model (b). Visualizations are based on accumulated local effect plots. Uncertainties for Hybrid-FINN (A) are based on Monte-Carlo Dropout. Swp is the observed soil water potential.}
\label{fig:figure_5}
\end{figure}

The inferred functional forms of the growth responses differ markedly between Hybrid-FINN and Process-FINN (inferred by accumulated local effect plots \citep{apley2020visualizing}). The DNN growth process in Hybrid-FINN  is not saturating and is mostly linear for all PFTs, whereas the relative growth of Process-FINN saturates (Fig.~\ref{fig:figure_5}). Additionally, the hybrid growth process exhibits growth with little or no available light for fast growers and short-lived breeders (Fig.~\ref{fig:figure_5}). Note that fast growers die quickly in deep shade (Fig.~S7), so the learned response is unlikely to occur in reality.

We also find that parameter estimates of other processes also differ between Hybrid-FINN and Process-FINN (Fig.~S7,~S8). Especially estimates for the mortality parameters of ‘short-lived breeders’ show opposite effects, whereas Hybrid-FINN captures the expected pattern with a strong positive effect of size (dbh) on mortality, mirroring the demographic trade-off of this PFT with small stature and high mortality rates \citep{rugerFastSlowContinuum2018}.

\section*{Discussion}

We found that replacing a mechanistic process with a DNN enabled FINN to identify the true functional form of that process in a simulated scenario, thereby improving its predictive performance. Using BCI data, we demonstrate that the hybrid-FINN model, in which growth is replaced by a DNN, improves predictive performance and simulates more realistic succession trajectories than the mechanistic FINN version (Process FINN) and a naïve DNN. Furthermore, we used explainable AI to interpret the hybrid-FINN and found that it had learned a plausible growth process. 

Our results showed that a DNN embedded within a DVM and joint calibration can reduce the structural model bias of the replaced process. From the beginning, the DNN, which replaces the growth process, is calibrated to predict emergent patterns of forest dynamics due to the DVM scaffold (Fig.~\ref{fig:figure_4}). Additionally, Process-FINN and Hybrid-FINN exhibit distinct parameter estimates for mortality and recruitment (see Fig.~S7,~S9), suggesting a compensatory bias in both models \citep{cameronIssuesCalibratingModels2022}. While it is unclear whether the compensatory bias in Hybrid-FINN weakened or merely shifted, these results demonstrate that joint calibration aligns the processes with each other. Hence, we hypothesize that an external plug-in approach would likely have shown worse predictive performance \citep[see also][]{hulsmannHowKillTree2018, cailleretBayesianCalibrationGrowthdependent2019}, but we did not test this directly. Lastly, FINN confirms that joint calibration allows processes to be calibrated without process-level observations (Fig.~S9).

When comparing Process-FINN, Hybrid-FINN, and the naïve DNN, all three models can be calibrated to provide accurate stand-level fits (basal area, stem density) but not accurate individual demographic processes. This is not surprising, as it is  in line with previous findings that DVMs of different complexity levels can predict similar forest dynamics \citep{bugmannTreeMortalitySubmodels2019,diaz-yanezTreeRegenerationModels2024,mahnkenAccuracyRealismGeneral2022}. The stand-level variables result from different processes within the model and thus suffer more from the problem of equifinality. Although we did not evaluate the models under long-term or extrapolative scenarios, simulations of succession trajectories showed reasonable patterns for Hybrid-FINN but not for Process-FINN. This confirms that that modest fit differences can amplify under long-term simulations. For this reason, we hypothesize that Hybrid-FINN would generalize better in even more extreme extrapolation tasks than that tested here.

From a DL perspective, Hybrid-FINN's substantial improvement over naïve DNNs emphasizes the value of a process scaffold: the surrounding DVM structure constrains the DNN's flexibility similarly to PINNs that encode constraints into the loss functions of DNNs \citep{wesselkampProcessInformedNeuralNetworks2024}, but makes the complexity more purposeful. Complexity is permitted only in processes specified as DNNs and within plausible patterns of forest dynamics. Although the naïve DNN was the worst model, its performance was better than expected. However, results showed implausible succession trajectories for the naive DNN (Fig.~S12). This is why we believe that more complex DNNs, such as recurrent neural networks (RNN), may also fail, as the surrounding process that constrains the model into producing emergent patterns of forest dynamics is missing. We could not test this theory because an RNN would likely overfit to the seven time-series data from the BCI censuses Nevertheless, these more complex DL approaches should be tested using more appropriate data in future research.

The learned growth function of Hybrid-FINN retained the ‘mechanistic’ size effect of Process-FINN but diverged sharply in its response to light: ‘fast growers’ and ‘short-lived breeders’ were predicted to grow even under deep shade, contradicting the classical assumption of no growth at zero light \citep{farquharBiochemicalModelPhotosynthetic1980}. This pattern aligns with field evidence that ‘short-lived breeders’ persist under closed canopies \citep{rugerFastSlowContinuum2018}. We discovered that the implausible result of light-demanding "fast growers" growing under deep shade is an interpretive extrapolation artifact. While the model does report an effect, this combination of zero light and growth for "fast growers" almost never occurs in the model (this also applies to Process-FINN). This explanation is supported by the estimated effect of shading on mortality (see Fig.~S7). Nevertheless, these artifacts stress that processes should only be considered within the emergent patterns of individual species, particularly for hybrid approaches, where DNN behavior in extrapolation regions is less predictable than that of deterministic functions. This also shows that further research is required to identify the limitations of this hybrid approach and develop potential solutions. For example, we refrained form including prior information in the DNN of the Hybrid-FINN but we could incorporate prior knowledge into the DNN via PINN to more precisely control its flexibility or determine its behavior in extreme extrapolation scenarios.

\subsection*{Conclusion}
With FINN, we introduce a flexible hybrid modeling approach that unites empirical and process-based modeling, offering the best of both worlds \citep{breimanStatisticalModelingTwo2001, hartigConnectingDynamicVegetation2012}. The DVM, informed by our process knowledge, serves as a constraining scaffold that controls the DNN's complexity as needed. Although we demonstrated that the FINN approach can achieve superior predictive performance and identify the functional form of processes, caution is advised. First, it is unclear whether all functional forms are identifiable at all: absence of species may be caused by high mortality or low growth. This makes it difficult to learn the full functional range of the functional forms, especially in low-support regions of the covariate space. Second, combining process knowledge and empirical data requires rethinking how we implement process knowledge and what kind of process knowledge is compatible with empirical data. DL allows to integrate complex data, such as remote sensing data or optical images, but it is unclear where they should enter the DVM. Overall, FINN provides a practical path to end-to-end differentiable calibration and empirical process discovery within the DVM that improves predictive performance while yielding interpretable hypotheses about processes of forest dynamics.

\section*{Methods}
This section details how FINN was calibrated based on simulated data and BCI forest inventory data. First, the concept of replacing a process with a deep neural network (DNN) is introduced using the example of the growth process. Second, the data preparation steps are described, including validation, variable calculation, and spatial aggregation based on the BCI plot. Third, the model training procedure is described, which is based on a joint loss across ecological responses and optimized using stochastic gradient descent. Finally, the validation and uncertainty estimation methods are presented, including spatial cross-validation and Monte Carlo dropout. A detailed model description is given in the Supplementary Information.

\subsection*{Hybrid modeling with Neural Networks}
If the process is set to hybrid modeling, it is replaced by a deep neural network (DNN) with two hidden layers, each containing 50 nodes and using ReLU activation functions with a dropout rate of 20\%. The DNN processes each cohort individually, characterized by diameter at breast height (dbh), light availability, species, number of trees, growth (for mortality), and site-specific environmental predictors. To improve convergence, dbh is divided by 100, and tree variables are log-transformed before being passed to the network. Species are encoded as integers and mapped into a low-dimensional space using an embedding layer before being concatenated with the other status variables, allowing the DNN to learn species-specific process functions. We used an embedding dimension of two.

In the hybrid mortality process, the process function simplifies to:
\begin{align*}
\pi_{i,p,c} = f(\text{dbh}_{i,p,c}, g_{i,p,c}, \text{trees}_{i,p,c}, \text{light}_{i,p,c}, \\ \text{species}_{i,p,c}, \text{env}_{i^{\text{(mortality)}},s})
\end{align*}

And in the hybrid growth process, the process function simplifies to:
\begin{align*}
g_{i,p,c} = f(\text{dbh}_{i,p,c}, \text{trees}_{i,p,c}, \text{light}_{i,p,c}, \\ \text{species}_{i,p,c}, \text{env}_{i^{\text{(growth)}},s})
\end{align*}

Here, \(f\) corresponds to the DNN, which is intended to approximate the mortality or growth process from the data. In the mortality DNN, the final activation is a sigmoid function serving as an inverse link function to output probabilities. For the growth DNN, the last activation function takes the form:
\begin{align*}
y = e^{x - e}
\end{align*}
which we found to have good convergence properties.

\subsection*{Calibration data preparation}
The 50-ha BCI plot is located at $9.15^{\circ}$ N, $79.85^{\circ}$ W in the Panama canal and covers a tropical primary forest with high tree species diversity. Here we focus on seven forest inventories collected between 1985 and 2015, which represent one initial state and six subsequent time periods for model calibration.

We calculated six stand variables that served as input to the joint likelihood explained below. Before calculating the species-specific variables, we cleaned the raw data by checking the plausibility of dbh measurements by evaluating the difference of measured dbh between inventories. Only changes that fall within the range of an absolute dbh change from -0.2 and 4.5 cm or a relative dbh change from -0.01 and 5 were considered valid. All other dbh were set to NA and excluded for the calculation of basal area, mean dbh, and growth. In the case of invalid dbh where the previous and following dbh were plausible, missing values were interpolated.

The observed variables for each timestep $t$, site $i$, patch $p$, and species $s$ are defined as the response vector $y_{t,i,p,s}$ and consist of basal area ($ba$, m\textsuperscript{2} per patch), number of trees ($trees$, N per patch), average diameter at breast height ($dbh$, cm), average relative growth rate ($g$, \%/100), mortality rate ($m$, \%/100), and the regeneration rate ($r$, N/ha):

\begin{align*}
ba_{t,i,p,s} &= \sum_{k=1}^{n} \left( \frac{\pi \cdot (dbh_{t,i,p,s,k} \cdot 0.01)^2}{4} \right) \\
trees_{t,i,p,s}^{(alive)} &= \sum_{k=1}^{n} trees_{t,i,p,s,k}^{(alive)} \\
dbh_{t,i,p,s} &= \frac{\sum_{k=1}^{n} dbh_{t,i,p,s,k}}{n} \\
g_{t,i,p,s} &= \left( \frac{dbh_{t,i,p,s}}{dbh_{t-1,i,p,s}} \right)^{1/\Delta t} - 1 \\
m_{t,i,p,s} &= \left( \frac{trees_{t,i,p,s}^{(alive)} - trees_{t,i,p,s}^{(dead)}}{trees_{t-1,i,p,s}^{(alive)}} \right)^{1/\Delta t} - 1 \\
r_{t,i,p,s} &= \sum_{k=1}^{n} trees_{t,i,p,s,k}^{(new)}
\end{align*}

In all equations k denotes individual cohorts or trees. All responses are finally aggregated per site $i$ by averaging over all patches $p$:
\begin{align*}
y_{t,i,s} = \frac{\sum_{p=1}^{n} y_{t,i,p,s}}{n}
\end{align*}
which results in the response used for calculating the joint likelihood.

After joining the published data from the BCI forest plot \citep{conditDemographicTrendsClimate2017,condit_bci_2019,condit_census_2019,condit_complete_2019} and the list of tree and shrub species from \citealp{rugerDemographicTradeoffsPredict2020}, a total of 281 tree and shrub species were left; these were aggregated into 145 genera or assigned to the five plant functional types (PFTs) described by \citealp{rugerFastSlowContinuum2018}. The five PFTs facilitate the interpretation with respect to key demographic trade-offs (i.e., ecological plausibility) and also allow validation with the predictive performance of the data-driven but much simpler PPA model \citep{rugerDemographicTradeoffsPredict2020}. To calibrate FINN as a simple process model, PFTs require significantly fewer parameters ($N=80$) than a genus-based approach ($N=2320$), yet the genus-based approach involves fewer assumptions regarding species differences and demographic trade-offs. The latter, however, introduces greater complexity, complicates interpretation, and demands substantially more computational resources. We calibrated the model using both variants but only provide a detailed analysis of the PFT-based model.

Climate data was derived from the hourly time series of climate variables published by \citep{faybishenko_pa-bci_2021}, where we calculated daily values for the precipitation sum, sum of solar radiation, average relative humidity, and average daily temperature. These were further aggregated to annual values by summing daily precipitation ($Prec$) and solar radiation ($SR\_kW\_m2$), and computing the annual minimum of daily temperature ($T\_min$) and maximum of daily temperature ($T\_max$) and mean relative humidity ($RH\_prc$).

\begin{table}[t!]
\centering
\caption{Climatic variables aggregated as averages for each observation period.}
\begin{tabular}{lrrrrr}
Year & Prec & SR\_kW\_m2 & RH\_prc & T\_max & T\_min \\
\midrule
1985 & 1638.05 & 1570.86 & 88.80 & 27.38 & 22.44 \\
1990 & 1995.19 & 1580.57 & 89.18 & 27.38 & 22.61 \\
1995 & 2308.96 & 1630.16 & 89.42 & 27.56 & 22.69 \\
2000 & 2399.69 & 1654.88 & 90.40 & 29.11 & 23.60 \\
2005 & 2141.88 & 1560.83 & 90.54 & 28.00 & 23.06 \\
2010 & 2411.12 & 1555.84 & 89.68 & 27.76 & 22.48 \\
2015 & 2347.47 & 1576.47 & 90.69 & 28.07 & 23.43 \\
\bottomrule
\end{tabular}
\label{tab:table_1}
\end{table}

Soil water potential (swp) was derived from \citep{kupersDrySeasonSoil2019}. We tested the spatial variation of all available variables for site water potential and decided to choose the site water potential for the late dry season because it had the highest spatial variation.

Spatial aggregation was done by separating the 50-ha plot into 500 rectangles (40$\times$25 m) with an area of 1000 m\textsuperscript{2}, each representing a forest patch. These 500 patches were used as 500 individual sites with unique swp for each of the 500 patches, as well as 20 sites each containing 25 patches and the average swp over these 25 patches. This allowed us to test two trade-offs for model calibration with regard to the spatial representation (Fig.~S9).

We also tested two variants for the temporal resolution. First, at the resolution of the original data with a 5-year time step, where climatic variables were averaged for each time period. Second, at an annual time step with $\Delta t = 5$ (Fig.~S9).

The five-fold spatially blocked cross-validation is based on five folds that were identified by randomly assigning spatial blocks – defined using the estimated spatial autocorrelation range derived from a fitted variogram model of the soil water potential (SWP) data – to folds using the \texttt{cv\_spatial()} function from the \texttt{blockCV} package.

\subsection*{Model training}
We used a joint loss function for the different responses. We used mean squared error for dbh, ba, mortality and growth rates, and negative binomial likelihoods for the number of trees and recruits. Interim results showed that convergence benefits from unequally weighting the different losses. Average dbh was weighted down (to 1/10), and, in the case of hybrid modeling, average growth was given a higher weight (10.0); all other responses were given a weight of 1.0.

The models were trained for 8,000 epochs with a batch size of 50\%. The learning rate of the optimizer was set to 0.01. NA in the responses were masked in the joint loss function. For optimization, we used ADAM, a popular first-order optimizer from deep learning that can train DL models with millions of parameters. We refrained from using more sophisticated optimizers, such as second-order optimizers, because they scale poorly with the number of parameters. First-order optimizers from deep learning combine several principles of optimization theory, such as momentum and adaptive learning rates, that increase their reliability enormously. We demonstrated that these "simple" optimizers can efficiently inverse calibrate FINN (Fig.~S10), offering a clear runtime advantage over other approaches such as MCMC sampling (Fig.~S11).

\subsection*{Hybrid Modeling of tree growth improves predictions of forest dynamics}
We divided the BCI plot into five spatial blocks for a five-fold spatially blocked cross-validation \citep{robertsCrossvalidationStrategiesData2017}, which provides accurate estimates of the model’s generalizability. We averaged the predictions for the five spatial holdouts and calculated the Spearman correlation factors between predicted and observed responses (Fig.~\ref{fig:figure_3}). We compared basal area (ba), number of trees, diameter at breast height (dbh), regeneration (ingrowth), growth, and mortality rates. Although our model is stochastic with respect to mortality and regeneration, the site-specific aggregation of the output improves optimization stability.

Uncertainties of the Hybrid-FINN for the growth predictions are based on Monte Carlo Dropout. The dropout rate was set to $( p = 10\% )$ during training. After the training, dropout was not turned off, and we averaged and calculated the standard deviation of 100 ALE predictions from Hybrid-FINN.

\section*{Data, Materials, and Software Availability}
The original BCI plot data is available from \url{https://doi.org/10.15146/5xcp-0d46}, meteorological data from \url{https://doi.org/10.15486/ngt/1771850}, water potential map from \url{https://doi.org/10.6084/m9.figshare.c.4372898.v1}, and species list from \url{https://doi.org/10.1126/science.aaz4797}. Scripts and code to reproduce the analysis and figures are available from \url{https://github.com/FINNverse/FINNetAl}. The FINN model documentation is provided in the SI Appendix. FINN is available as R package and can be installed from \url{https://github.com/FINNverse/FINN}.

\section*{Acknowledgements}
We are grateful for the valuable comments and insightful discussions with Florian Hartig, Carsten Dormann, Harald Bugmann, Lisa Hülsmann, Nadja Rüger, and Jonas Stillhard, as well as the support from the YOMOS AG of the Ecological Society of Germany, Austria, and Switzerland (GFÖ).

\section*{Funding Information}
Yannek Käber received funding from European Union Horizon Scheme (project Wildcard no. 101081177).

\newpage

\FloatBarrier

\bibliographystyle{unsrtnat}  

\appendix

\section*{Appendix}
\renewcommand{\thefigure}{S\arabic{figure}}
\setcounter{figure}{0}
\renewcommand{\thetable}{S\arabic{table}}
\setcounter{table}{0}
\renewcommand{\theequation}{S\arabic{equation}}
\setcounter{equation}{0}

The supporting material includes a detailed description of the FINN model and additional results from simulation exercises conducted to generate more insight into the applied methods.

\section*{Model description}
FINN simulates forest dynamics with the demographic processes of regeneration, growth, and mortality along with competition and environmental inputs in discrete timesteps. To integrate process knowledge of individual processes with observational data, FINN provides a basic structure that allows defining mechanistic processes, integrating data models, or a combination of both (Figure 1, main text). The temporal and spatial resolution of the model is determined by the process definitions and its inputs. To integrate process knowledge of individual processes with observational data, FINN provides a basic structure that allows defining mechanistic processes, integrating data models, or a combination of both (Figure 1). The temporal and spatial resolution of the model is determined by the process definitions and its inputs. In the following the basic structure of FINN and its components is explained, along with one variant of a predetermined model structure (i.e., mechanistic process).

\subsection*{Cohorts}
Individual trees are represented in cohorts. Each cohort is defined by the diameter at breast height of the tree (dbh), the number of trees (trees), and the species identity as an integer (species). At each timestep t these properties together define a cohort (C) and are represented as arrays of equal dimensions for the number of sites (i), patches, (p), and cohorts (c).

\begin{equation}
C_{i,p,c} =
\bigl(
  \text{dbh}_{i,p,c},
  \text{trees}_{i,p,c},
  \text{species}_{i,p,c},
  \text{cohortID}_{i,p,c}
\bigr),
\qquad
c \in \{1,\dots,C_t\}
\end{equation}

The species array is implemented as such but only used for indexing species-specific parameters. In the process equations described here it is always abbreviated with s to simplify the equations below.
Due to regeneration and mortality, the number of cohorts changes with each timestep t, which leads to different indices c for each timestep. Dead cohorts are removed from the arrays for computational reasons. To keep track of the individual cohorts another array stores the cohort ID of each individual cohort.

\begin{equation}
C_{t+1} = \text{FINN}\bigl(C_t,\,\text{env}_t,\,\theta_j\bigr)
\end{equation}

Updating cohorts
FINN simulates forest dynamics by updating the status of all cohorts in discrete timesteps t. Within this subsection the index t is used to explain the temporal mode in which the model operates.
$C_{t+1}=\text{FINN}\left(C_t,\text{env}_t,\theta_j\right)$
where $C_t$ are the cohorts at timestep t and $C_{t+1}$ are the cohorts after all demographic processes took place. Model parameters are denoted with $\theta_j$. The demographic processes take place in the following order.

\begin{align}
\text{light}_t^{(\text{growth})} &=
  \text{competition}\bigl(\text{dbh}_t,\text{trees}_t,\theta_j\bigr) \\[4pt]
g_t &=
  \text{growth}\bigl(\text{env}_t,\text{dbh}_t,
                     \text{light}_t^{(\text{growth})},\theta_j\bigr) \\[4pt]
\text{dbh}_{t+1} &=
  \text{dbh}_t + \text{dbh}_t \, g_t \\[4pt]
\text{light}_t^{(\text{mortality})} &=
  \text{competition}\bigl(\text{dbh}_{t+1},\text{trees}_t,\theta_j\bigr) \\[4pt]
m_t &=
  \text{mortality}\bigl(\text{env}_t,\text{dbh}_t,
                        \text{light}_t^{(\text{growth})},\theta_j\bigr) \\[4pt]
\text{trees}_{t+1}^{(\text{after mort.})} &=
  \text{trees}_t - m_t \\[4pt]
\text{light}_t^{(\text{regen})} &=
  \text{competition}\bigl(\text{dbh}_{t+1},
    \text{trees}_{t+1}^{(\text{after mort.})},\theta_j\bigr) \\[4pt]
r_t &=
  \text{regeneration}\bigl(\text{env}_t,
    \text{light}_t^{(\text{regen})},\theta_j\bigr) \\[4pt]
\text{trees}_{t+1} &=
  \text{trees}_{t+1}^{(\text{after mort.})}+r_t
\end{align}

After all processes took place, all cohort arrays dbh, trees, and cohortID are updated to remove dead cohorts and add new cohorts with the predefined dbh threshold at which trees regenerate. The chosen sequence of processes allows to simulate growth-dependent mortality, and implies that regeneration will only take place at the end of the simulated timestep. This, of course, has stronger implications for scenarios in which the timestep is large. However, there might be cases in which a different sequence or the approach of only updating the cohorts after all processes were computed is more suitable. Each individual process is explained in detail below. Note that subsequently timestep t is not included in the equations to avoid confusion with notations representing arrays with spatial dimensions site i, patch p, and cohort c. Unless indicated otherwise, in the following each equation is defined for each timestep t.

\subsection*{Competition}

FINN simulates forest dynamics in the following order. Competition simulates the interaction of trees based on their stature and species properties. It calculates the available light as a function of dbh, trees and species-specific parameters.

Basal area (ba) ($m^2 ha^{-1}$) serves as central currency for simulating competition among trees (Fig.~S1a). It is calculated for each cohort c at site s and patch p with
\begin{equation}
{\text{ba}}_{i,p,c}\ =\ \frac{\pi\ \cdot\ \left(\ \frac{{\text{dbh}}_{\text{i,p,c}}}{100\ \cdot\ 2}\ \right)^2\cdot\text{tree}\text{s}_{\text{i,p,c}}}{\text{patchsize}}
\end{equation}
where \(\text{db}\text{h}_{\text{i,p,c}}\) and \({\rm trees}_{\text{i,p,c}}\) are arrays representing the dbh in cm and number of trees, respectively, of cohort c in site i and patch p. ‘patchsize’ is the size of the patch in hectares. Using ba alone to quantify competitive pressure ignores species’ differences in competitive strength. To account for this, we compute a weighted basal area (\({\rm ba}_{i,p,c}^{\left(\text{weighted}\right)}\)) as proxy for the competitive pressure:
\begin{equation}
{\rm ba}_{i,p,c}^{\left(\text{weighted}\right)}={\text{ba}}_{i,p,c}\cdot{\rm compStr}_{\text{s}_{i,p,c}}\cdot0.1
\end{equation}
where compStr is an empirical parameter describing the shading efficiency of species s (Fig.~S1c,S1d). The factor 0.1 rescales the product so that ba remains on the same numerical order ($0-50\ m^2 ha^{-1}$) as other model inputs; it does not alter rank order or relative strength. Each cohort’s height is calculated via an allometric function:
\begin{equation}
h_{i,p,c}=\left(e^{\left(\frac{{\text{dbh}}_{\text{i,p,c}}\ {\text{parHeight}}_{\text{s}_{i,p,c}}}{{\text{dbh}}_{\text{i,p,c}}\ +\ 100}\right)}-1\right)\times100+0.001
\end{equation}

The allometric function was defined to cover a large range of linear height~dbh relationships (Fig.~S1b) without unrealistically high values (i.e. height > 120 m) with parHeight parameter values from 0 to 1.

Next, the height difference \(\Delta h\) for each cohort c to all other cohorts j is computed with
\begin{equation}
{\Delta h}_{i,p,c_j}=\ h_{i,p,c}-h_{i,p,c_j}-0.1,
\end{equation}
where \(h_{i,p,c_j}\) is a permuted version of \(h_{i,p,c}\) with the index j for each pairwise difference between cohorts c and j. For each pairwise difference a margin of 0.1 m is subtracted which ensures that a tree must be at least 0.1 m taller to have competitive advantages over neighboring trees. The value 0.1 m was chosen to overcome the issue of unrealistically high competition between small trees of similar size. The scaled height difference is multiplied by 100 and passed through a sigmoid function
\begin{equation}
\sigma\left(z\right)\ =\ \frac{1}{1\ +\ e^{-z}}
\end{equation}
to transform negative and positive height differences to 0 and 1, respectively. Or in simple terms: the transformation simplifies the height difference to the information of whether a neighboring tree is smaller (0) or taller (1). The output of this transformation serves as a mask to calculate the shading (i.e., competitive pressure) for each cohort c experienced from all cohorts j that are taller.
\begin{equation}
{\text{shade}}_{\text{i,p,c}}=\sum_{j=0}^{n}{{\rm ba}_{i,p,c_j}^{\left(\text{weighted}\right)}\ \bullet\ \sigma\!\left({\Delta h}_{i,p,c_j}\cdot100\right)}
\end{equation}
The sum over j then yields the accumulated basal area of all cohorts j that are taller than c (Fig.~S1c,S1d). Finally, available light for each cohort c on each patch p and each site i is given by
\begin{equation}
{\text{light}}_{\text{i,p,c}}=1-{\text{shade}}_{\text{i,p,c}}
\end{equation}
which then serves as input to the demographic processes.

\subsection*{Environment}
In its basic configuration FINN simulates the effect of environmental predictors through a neural network, which predicts how environmental variables regulate the process. The environmental effect, in the simplest case, is modeled as a linear function of environmental predictors. The environmental effect can be represented as:
\begin{equation}
\text{env}_{i,s}=%
\sigma\!\left(
  \beta_{s,0}
  +\beta_{s,1}\,x_{1,i}
  +\beta_{s,2}\,x_{2,i}
  +\cdots
  +\beta_{s,n}\,x_{n,i}
\right)
\end{equation}
where $\text{env}_{i,s}$ is the intermediate predicted value for species $s$ and site $i$ from the linear combination of environmental variables $x_{n,i}$.  
$\beta_{s,0}$ is the species-specific intercept, and $\beta_{s,1}, \beta_{s,2}, \dots, \beta_{s,n}$ are the species-specific coefficients for each predictor $x_{n,i}$.  
The variables $x_{n,i}$ represent environmental factors such as temperature, precipitation, or soil moisture.  
A sigmoid scales $\text{env}_{i,s}$ to the range $[0,1]$ to facilitate its combination with other mechanistic process elements.  
Environmental effects are calculated separately for regeneration, growth, and mortality, denoted $\text{env}_{i,s}^{(\text{regeneration})}$, $\text{env}_{i,s}^{(\text{growth})}$, and $\text{env}_{i,s}^{(\text{mortality})}$.

\subsection*{Regeneration}
Regeneration is the number of new trees that establish in a forest patch.  
The regeneration rate $\lambda_{i,p,s}$ for species $s$ on site $i$ and patch $p$ is calculated from available light and the species-specific parameter $\text{regLight}_{s}$, which defines the fraction of light required for successful regeneration.  
\begin{equation}
\lambda_{i,p,s}=%
\frac{%
  \sigma\!\bigl(10\,(\text{light}_{i,p}-\text{regLight}_{s})\bigr)
  -\sigma\!\bigl(-10\,\text{regLight}_{s}\bigr)%
}{%
  1-\sigma\!\bigl(-10\,(1-\text{regLight}_{s})\bigr)%
}\;
\text{env}_{i,s}^{(\text{regeneration})}
\end{equation}
Here $\lambda_{i,p,s}$ is the expected number of regenerating trees (ha$^{-1}$), $\text{light}_{i,p}$ is the output of the competition function at the forest floor ($h_{i,p}=0$), and smaller $\text{regLight}_{s}$ values indicate shade-tolerant species.  
A rescaled sigmoid keeps the function differentiable for gradient-based learning.  
Finally, the actual number of new trees is drawn from a negative-binomial distribution:
\begin{equation}
r_{i,p,s}\sim\text{NBinom}\!\bigl(\lambda_{i,p,s},\,\phi\bigr)
\end{equation}

The dbh at which trees establish can be defined. For the BCI case study it was chosen to be 1.

\subsection*{Growth}
The growth process models the diameter increase of each cohort as a product of a size term, a light-response term, and environmental effects.  
\begin{equation}
\text{size}_{i,p,c}^{(\text{growth})}
  =\exp\!\bigl(-\,\text{growthSize}_{s_{i,p,c}}\!\cdot dbh_{i,p,c}\bigr)
\end{equation}
\begin{equation}
\text{light}_{i,p,c}^{(\text{growth})}
  =\frac{%
     \sigma\!\bigl(10\,(\text{light}_{i,p,c}-\text{growthLight}_{s_{i,p,c}})\bigr)
     -\sigma\!\bigl(10\,\text{growthLight}_{s_{i,p,c}}\bigr)%
   }{%
     \sigma\!\bigl(10\,(1-\text{growthLight}_{s_{i,p,c}})\bigr)
     -\sigma\!\bigl(10\,\text{growthLight}_{s_{i,p,c}}\bigr)%
   }
\end{equation}
\begin{equation}
\text{g}_{i,p,c}
  =\text{light}_{i,p,c}^{(\text{growth})}\,
   \cdot\text{size}_{i,p,c}^{(\text{growth})}\,
   \cdot\text{env}_{i,s}^{(\text{growth})}
\end{equation}
Size-dependent growth decreases with increasing $dbh$, while the rescaled sigmoid introduces a light threshold for maximum growth.  
The absolute diameter increment is $\text{g}_{i,p,c}\times dbh_{i,p,c}$.

\subsection*{Mortality}
Mortality depends on growth, light, size, and environmental effects, combined through a logistic link:
\begin{equation}
\pi_{i,p,c}=%
\sigma\!\Bigl(
  \text{mortGrowth}_{s_{i,p,c}}\!\cdot\text{g}_{i,p,c}\;+\;
  \text{mortLight}_{s_{i,p,c}}\!\cdot\text{light}_{i,p,c}\;+\;
  \text{mortSize}_{s_{i,p,c}}\!\cdot dbh_{i,p,c}\!\cdot 0.01\;+\;
  \text{env}_{i,s}^{(\text{mortality})}
\Bigr)
\end{equation}
\begin{equation}
m_{i,p,c}\sim\text{Binom}\!\bigl(\pi_{i,p,c}\bigr)
\end{equation}
Growth-dependent mortality couples the process to simulated growth, while size-dependent mortality is scaled by $0.01$ to keep parameters on comparable magnitudes.  
The binomial draw is performed for each tree in each cohort, yielding the number of dead trees.

\section*{Additional Results}
\subsection*{Inference of process parameters in a dynamic forest model with stochastic gradient descent}
After training Process-FINN on the BCI forest plot, we generated 1,000 simulations from the model and trained Process-FINN for each simulation in 4,000 iterations (epochs), which allowed us to assess whether the true parameters of the initially trained models can be recovered during calibration. 
We first tested whether gradient-based optimization can reliably calibrate parameters in a DVM, mirroring the classic inverse calibration method \citep{vanoijenBayesianCalibrationProcessbased2005}, a necessary condition for replacing processes with deep neural networks. Using simulations, we evaluated the reliability of gradient-based optimization for Process-FINN. Stochastic gradient descent successfully recovered parameters for all plant functional types (PFT) parameters and environmental niche effects for each process without bias (Fig.~S10). Our results align with findings from differentiable ODEs \citep{boussangePartitioningTimeSeries2024}. One exception is the estimate for the light response of regeneration, which we attribute to the fact that the parameters of a negative binomial distribution with high dispersion can only be recovered with very high sample sizes (see also \citealp{kaberInferringTreeRegeneration2024}). Despite this exception, our findings show that stochastic gradient descent is a viable method for inferring parameters of a process-based forest model. Similar to \citealp{boussangePartitioningTimeSeries2024}, interim results showed that accumulating gradients throughout the entire time series resulted presumably in vanishing gradients and high computational costs; hence, we clipped gradients after each observed response in time.

A challenge in inverse calibration of DVM is that observational data are often incomplete or imbalanced \citep{cameronIssuesCalibratingModels2022,oberprillerRobustStatisticalInference2021}. However, DVMs link multiple processes, where the output of one process can serve as an input for other processes. As a result, a DVM can achieve good predictions for processes even when they are not directly observed \citep{fisherVegetationDemographicsEarth2018}.

To understand this in FINN, we simulated the availability of different process observations by omitting specific observed variables. We find that, even calibrating only with basal area as a response, dbh and trees can be predicted well ($\rho > 0.7$), which is not surprising since basal area is calculated from DBH and the number of trees, and the relation of these variables is defined in the model. For demographic rates, predictions remained accurate as long as at least one rate was observed (Fig. S9). This suggests that the observation of only one demographic rate, combined with variables of stand structure, may provide sufficient information to inform the other two demographic processes.

\subsection*{Recovery of a functional form in Hybrid-FINN}
The results shown in Figure 2 in the main text were generated from the simulated dynamics of a single species with a known set of parameters to test whether a neural network can recover the true functional form of a process. The simulated dynamic used for the recovery is shown in Fig. S3. The simulation parameters were chosen so that a gradient of artificial environmental conditions across 20 sites shows a clear uniform niche pattern. We used 10 patches per site and simulated for a total of 500 timesteps. The calibration was done for years 10 to 20 in the initial establishment phase. The following parameters were chosen. 
For regeneration: $\text{intercept}_{\text{regeneration}} = 2$, $\text{env}_\text{regeneration} = 0$, $\text{regLight} = 0.9$;
for mortality: $\text{intercept}_{\text{mortality}} = -3$, $\text{env}_\text{mortality} = 0$, $\text{mortLight} = 0.3$, $\text{mortSize} = -3$, $\text{mortGrowth} = 0.5$;
for competition: $\text{compHeight} = 0.6$, $\text{compStr} = 0.5$.
For growth, the parameters were chosen only for the true model M1 with $\text{intercept}_{\text{growth}} = 0$, $\text{env}_\text{growth} = 1.3
$, $\text{growthLight} = 0.1$, $\text{growthSize} = 0.03$.

\subsection*{Emergent pattern}
We compared the emergent pattern between the observed and simulated dynamics in the BCI plot. Fig. S5 compares observations vs. predictions of Process-FINN and Hybrid-FINN over years. Fig. S6 shows the basic relationship between individual tree dbh and mortality and growth rates as well as the basal area increment (BAI). The patterns match well and show that the basic relationships between the observed and simulated data match. Note that slight differences exist, especially for large trees, where the overall variance (of differences) is larger.

\subsection*{Parameter estimates Hybrid-FINN vs. Process-FINN}
We compared the parameter estimates from Process-FINN and Hybrid-FINN. Both models were calibrated with the BCI data each representing a different strategy for model calibration. While the calibrated Process-FINN resembles the classic inverse calibration approach, Hybrid-FINN provides high flexibility for the calibration to find the functional form of the growth process. Fig. S7 shows the estimates for the basic model parameters, while Fig. S8 shows the estimates for the environmental response. These results illustrate a key finding of the study: too rigid assumptions putting too many constraints on a single process can affect parameter estimates of other processes. In our case this was most evident for the PFT of short-lived breeders (SLB), where the mechanistic growth process led too opposite effect sizes of the mortality effect.

\subsection*{Inverse calibration of unobserved processes}
Often data availability is a crucial limitation of model calibration. In theory dynamic models should allow for the calibration of latent (i.e., unobserved) processes. We tested this by calibrating the Process-FINN with different sets of response variables. First, only with basal area (ba) which is closely related to the stands total basal area. Second with all variables of stand structure (basal area, N of trees, and average dbh). Third, with all combination of observed demographic processes (regeneration, growth, mortality). We find that in most cases the observation of only on process was sufficient to achieve good calibration results (Fig. S9). Note that we show the performance of different temporal and spatial resolutions in Fig. S9. Each 'period35' refers to annual timesteps, 'period7' refers to 5 year intervals, '25patches' and '1patch' refers to the spatial aggregation (i.e., the number of patches) considered for a single site. Throughout the manuscript we focus on the variant with 5 year intervals and 25 patches ('period7' and '25patches').

\begin{figure}
\centering
\includegraphics[width=\textwidth]{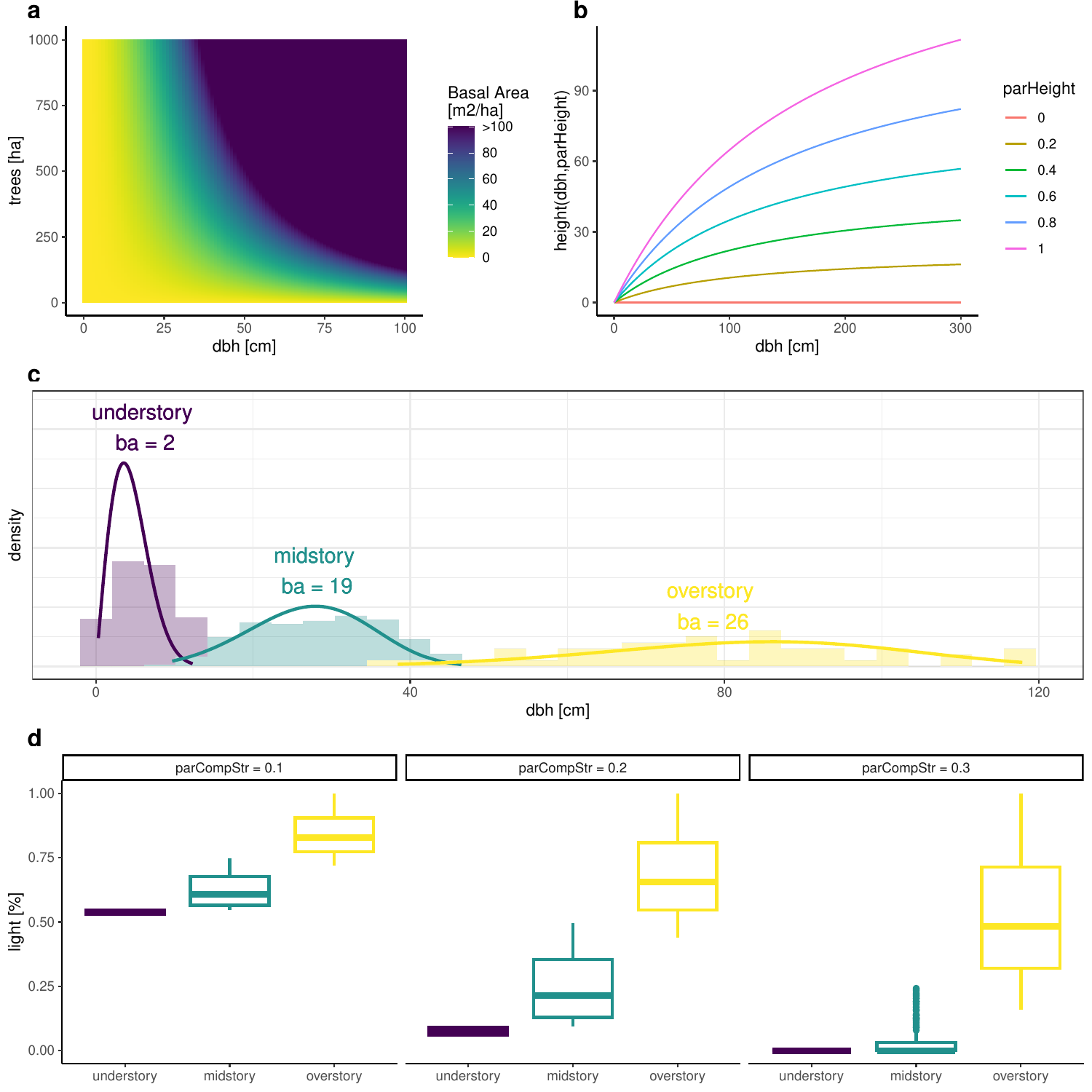}
\caption{Properties of the competition function. a) shows the relationship between dbh and trees; b) shows the allometric equation with which the height is derieved from the dbh. parHeight is a species specific parameter; c) shows randomly drawn dbh values from a Weibull distribution for three different sizes (i.e., canopy layers) to illustrate in d) how different layers receive different light with the species specific parameter for parCompStr set to the same values for all layers.}
\end{figure}

\begin{figure}
\centering
\includegraphics[width=\textwidth]{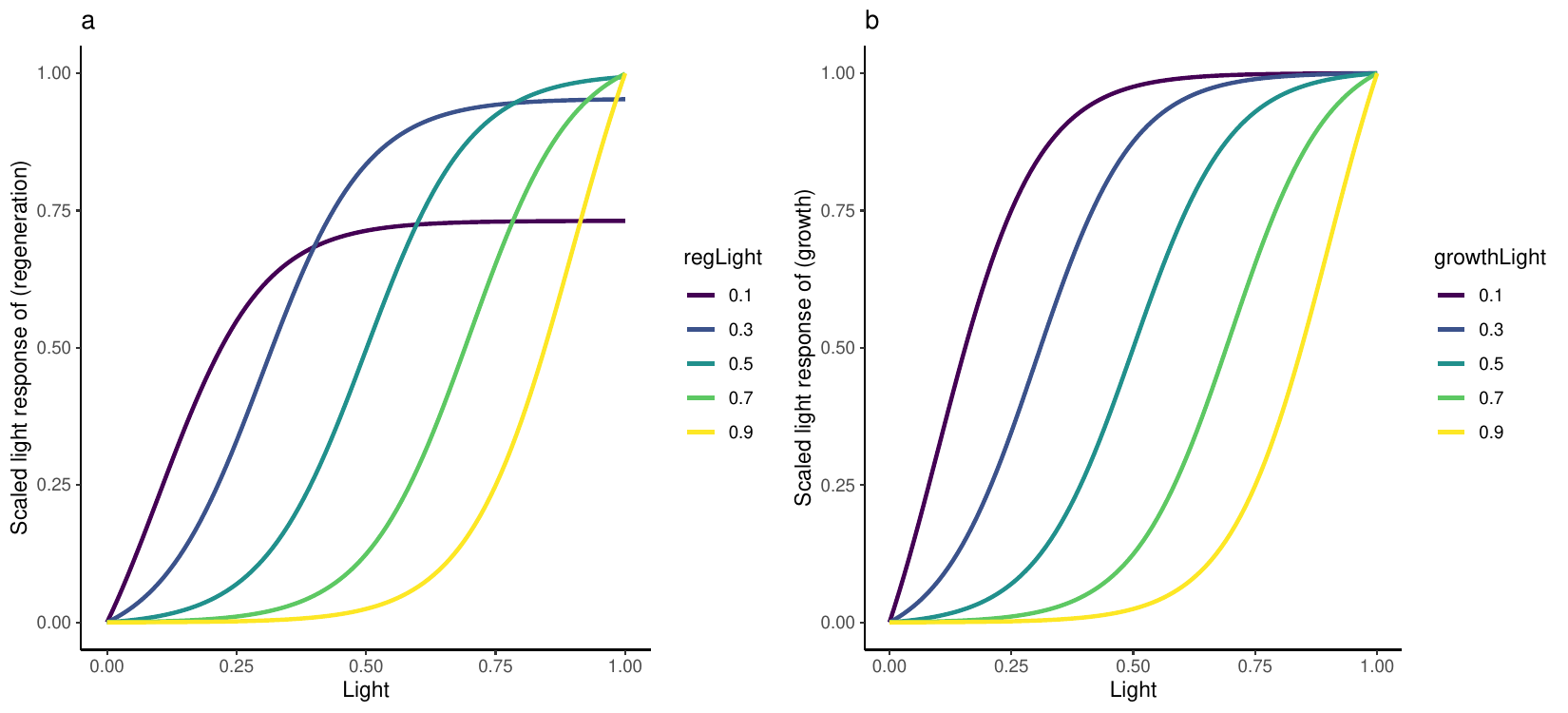}
\caption{Comparison of light response in Process-FINN for regeneration (a) and growth response (b)}
\end{figure}

\begin{figure}
\centering
\includegraphics[width=\textwidth]{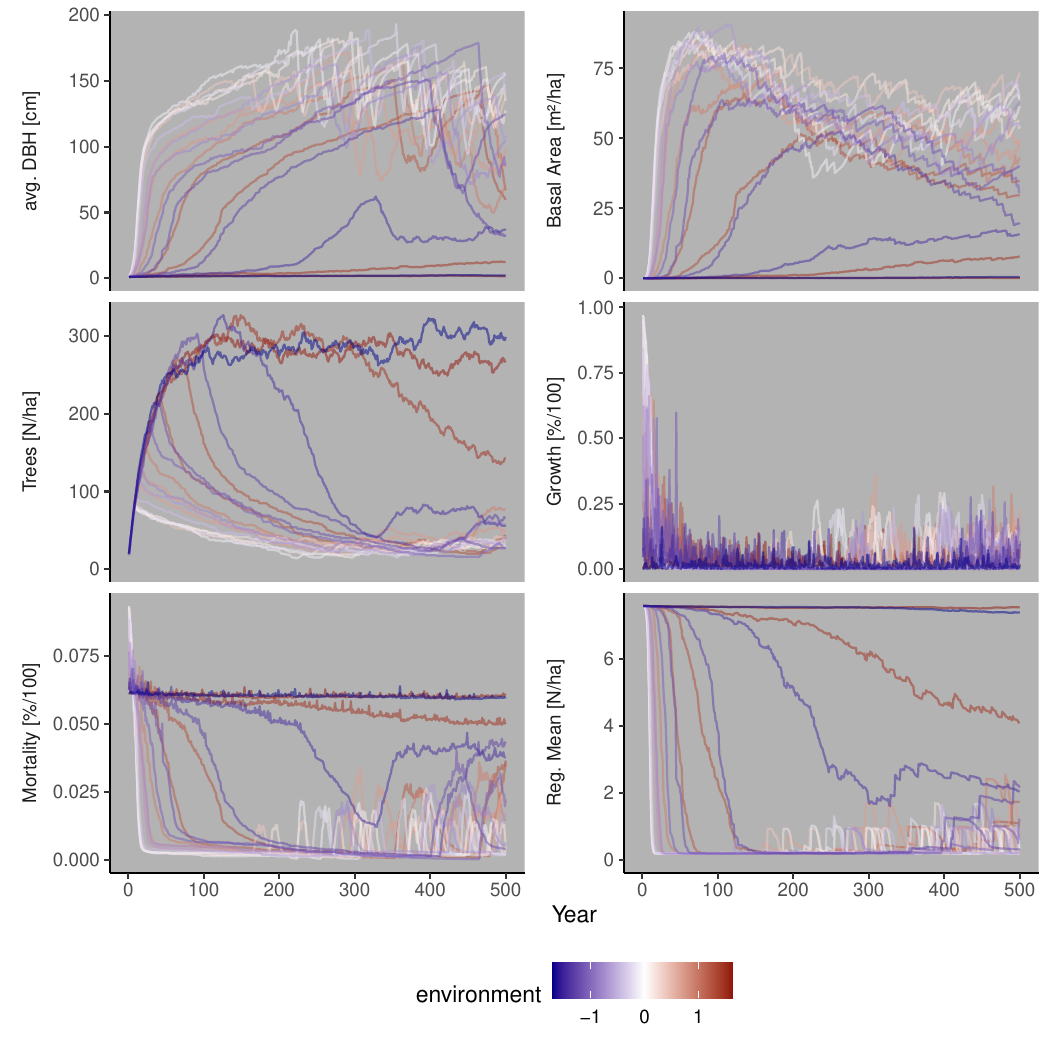}
\caption{Simulated succession of one species at 20 sites with different environmental conditions. Each panel shows the trajectory of forest stand variables and demographic rates. The color of each line indicates the average value for the environmental variable at each site}
\end{figure}

\begin{figure}
\centering
\includegraphics[width=\textwidth]{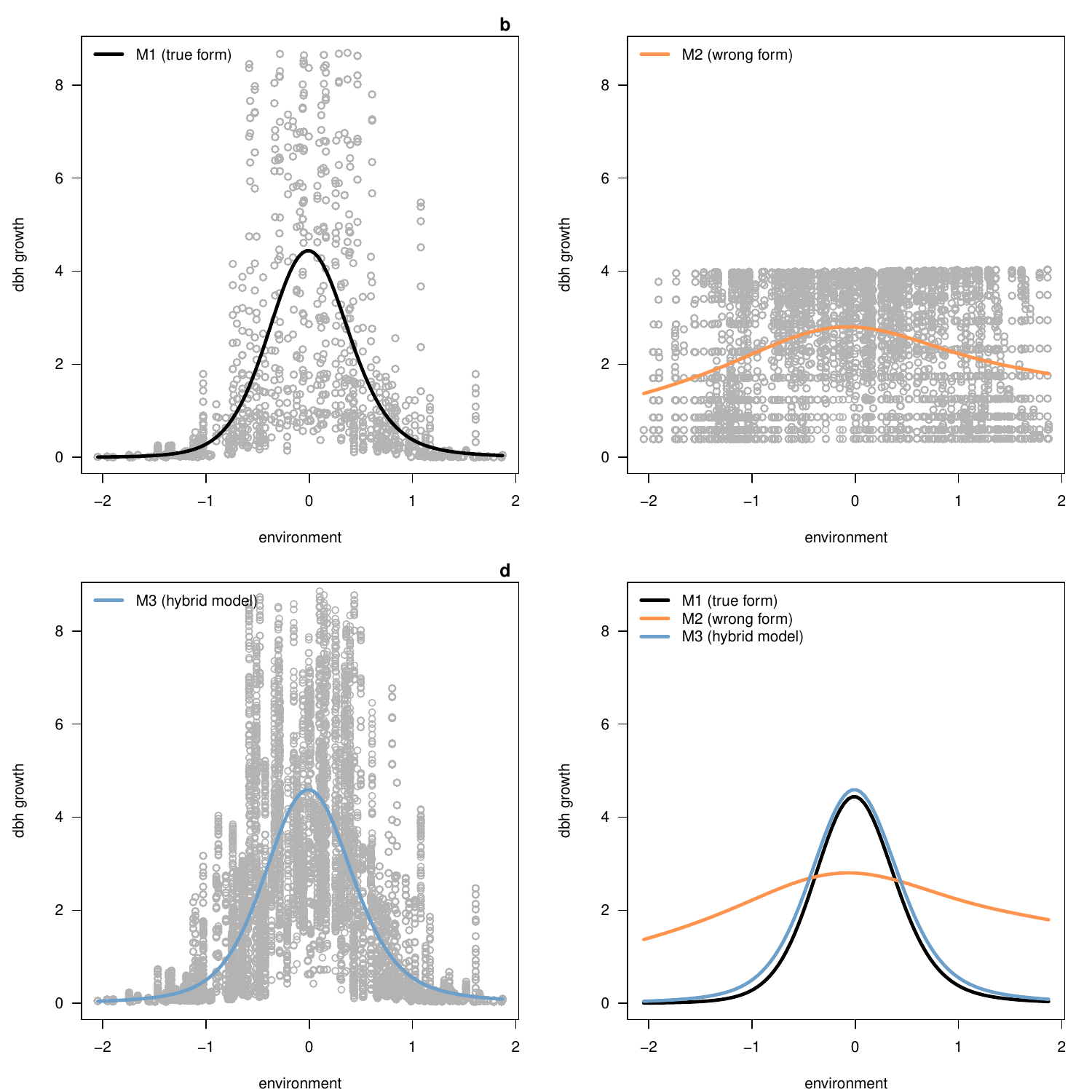}
\caption{Predicted dbh growth rate (y-axis) over environment (x-axis) from simulated succession of a single species with the true model M1 with a quadratic effect of environment (top-left), the wrong model M2 with a linear effect of environment on growth (top-right), the trained hybrid model (M3), and all predictions combined in one plot. The lines show fitted splines based on the formula $y \text{\textasciitilde} s(x, k = 5)$ with the \texttt{mgcv} package}
\end{figure}

\begin{figure}
\centering
\includegraphics[width=\textwidth]{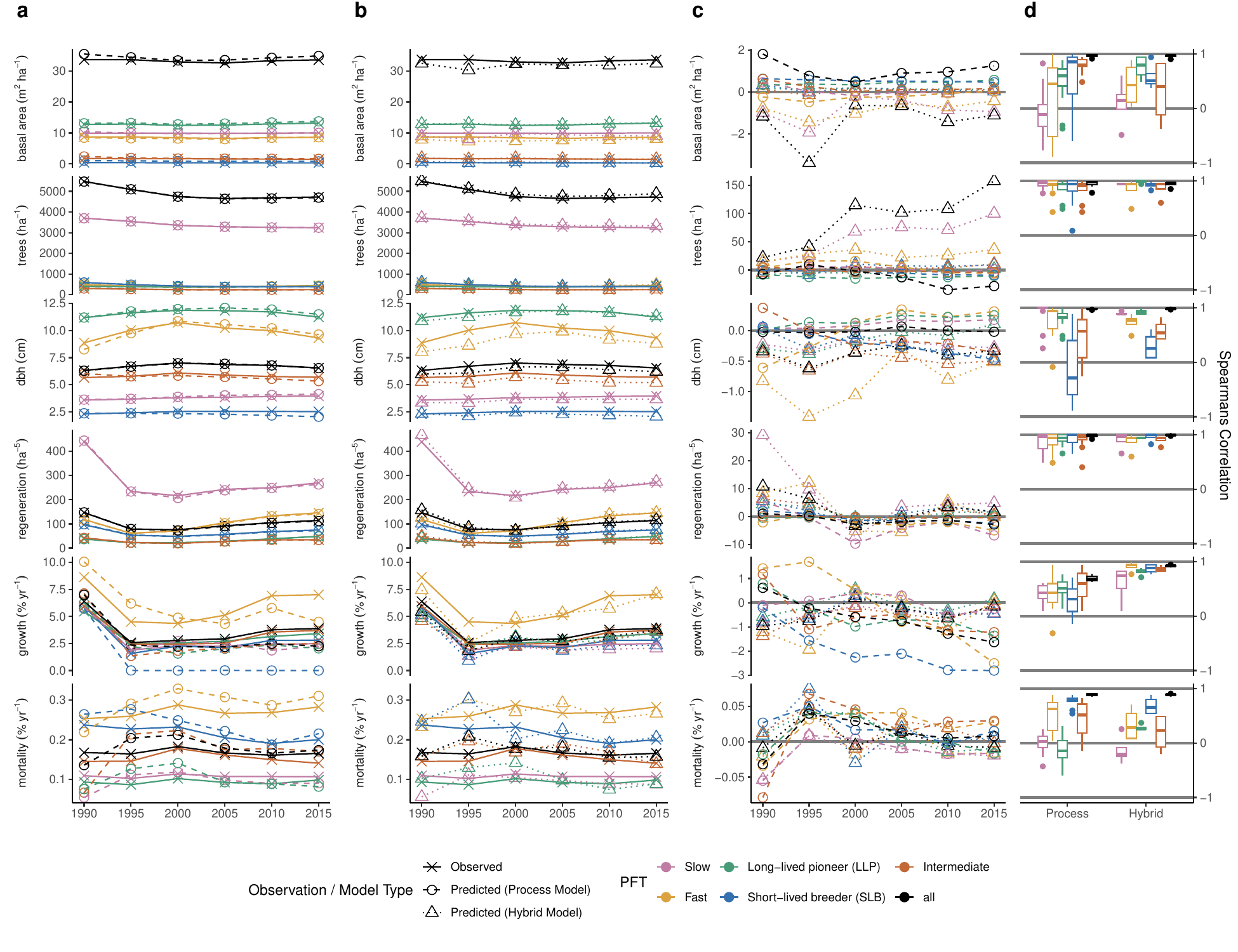}
\caption{Observed and predicted forest structure and demographic rates for five PFTs. (a) and (b) display the observed values (straight line \& crosses) vs.\ the predicted values from the process model (dashed line \& circles) and the hybrid model (dotted line \& triangles), respectively. (c) shows the absolute difference between the observed values and the predictions of the process and hybrid model. (d) shows the calculated Spearman correlation from the five-fold blocked spatial cross-validation. The first two rows show the observed and predicted values for basal area and tree density. The third and fourth row show the observed and predicted values for diameter at breast height (dbh), regeneration, growth, and mortality}
\end{figure}

\begin{figure}
\centering
\includegraphics[width=\textwidth]{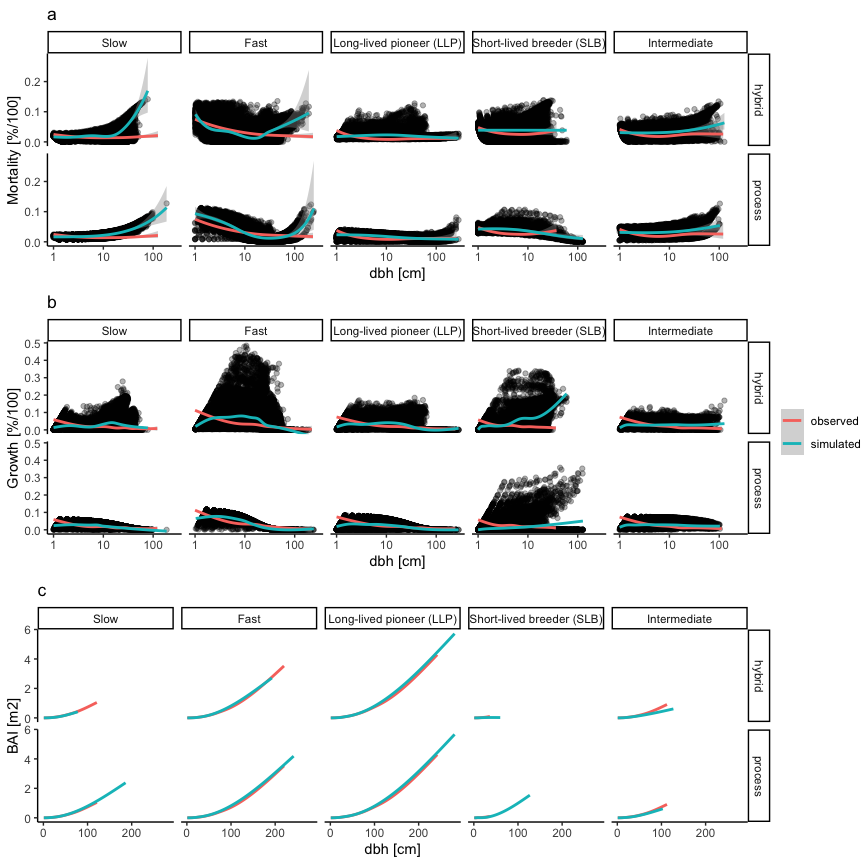}
\caption{Annual growth and mortality rates of trees vs.\ dbh. Observed values (red line) and simulated values from the process model and the hybrid model (cyan line). The lines show fitted GAMs to the data and visualize how well the simulated patterns match the observed patterns. Note that the observed data only contain 0 and 1 values for mortality while the simulated values show the estimated mortality probability. Black dots show the observed values. Note that all rates represent changes within 5 year intervals and are therefor much higher than annual rates.}
\end{figure}

\begin{figure}
\centering
\includegraphics[width=\textwidth]{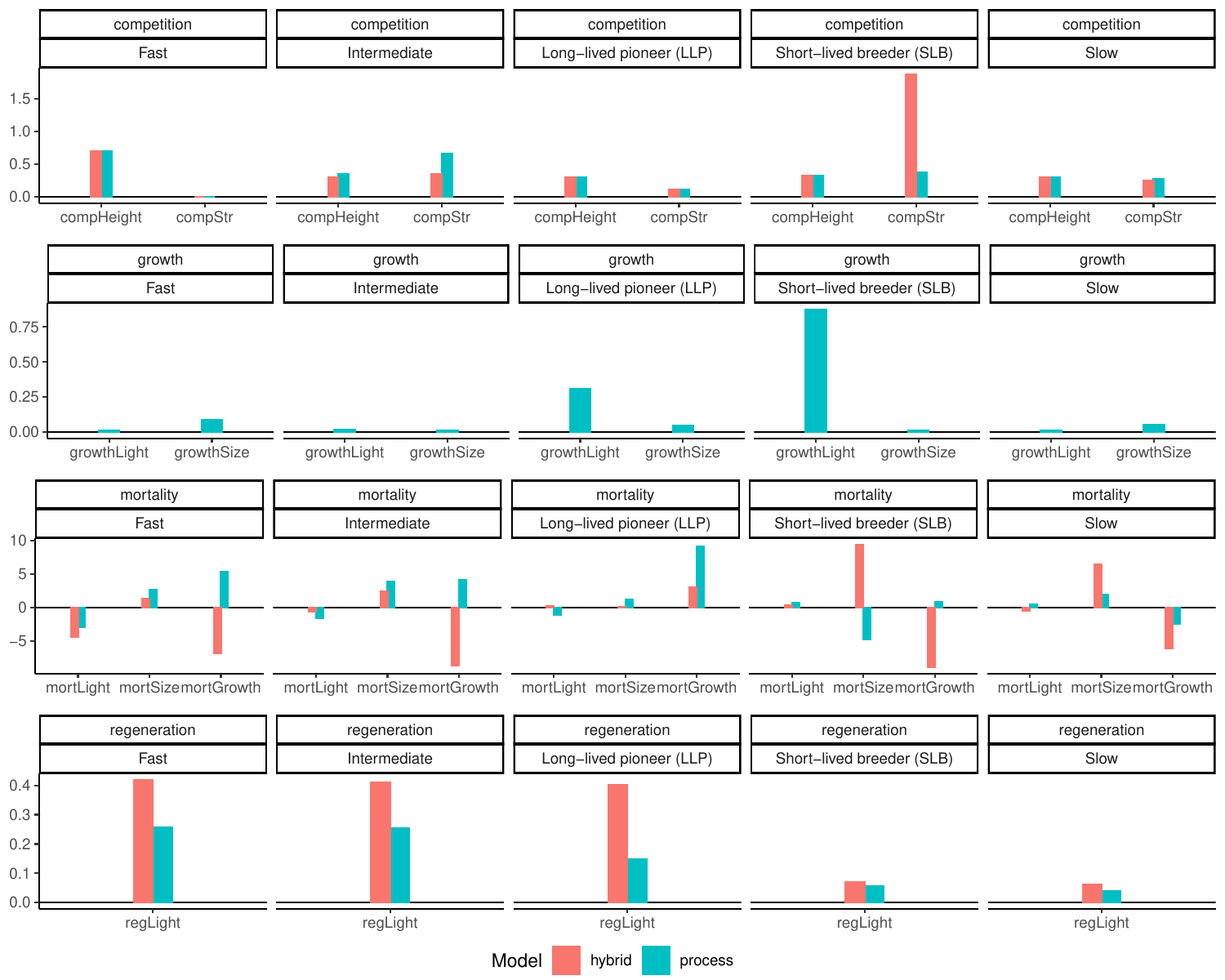}
\caption{Estimated process parameters for Process-FINN and Hybrid-FINN calibrated on the BCI forest inventory data}
\end{figure}

\begin{figure}
\centering
\includegraphics[width=\textwidth]{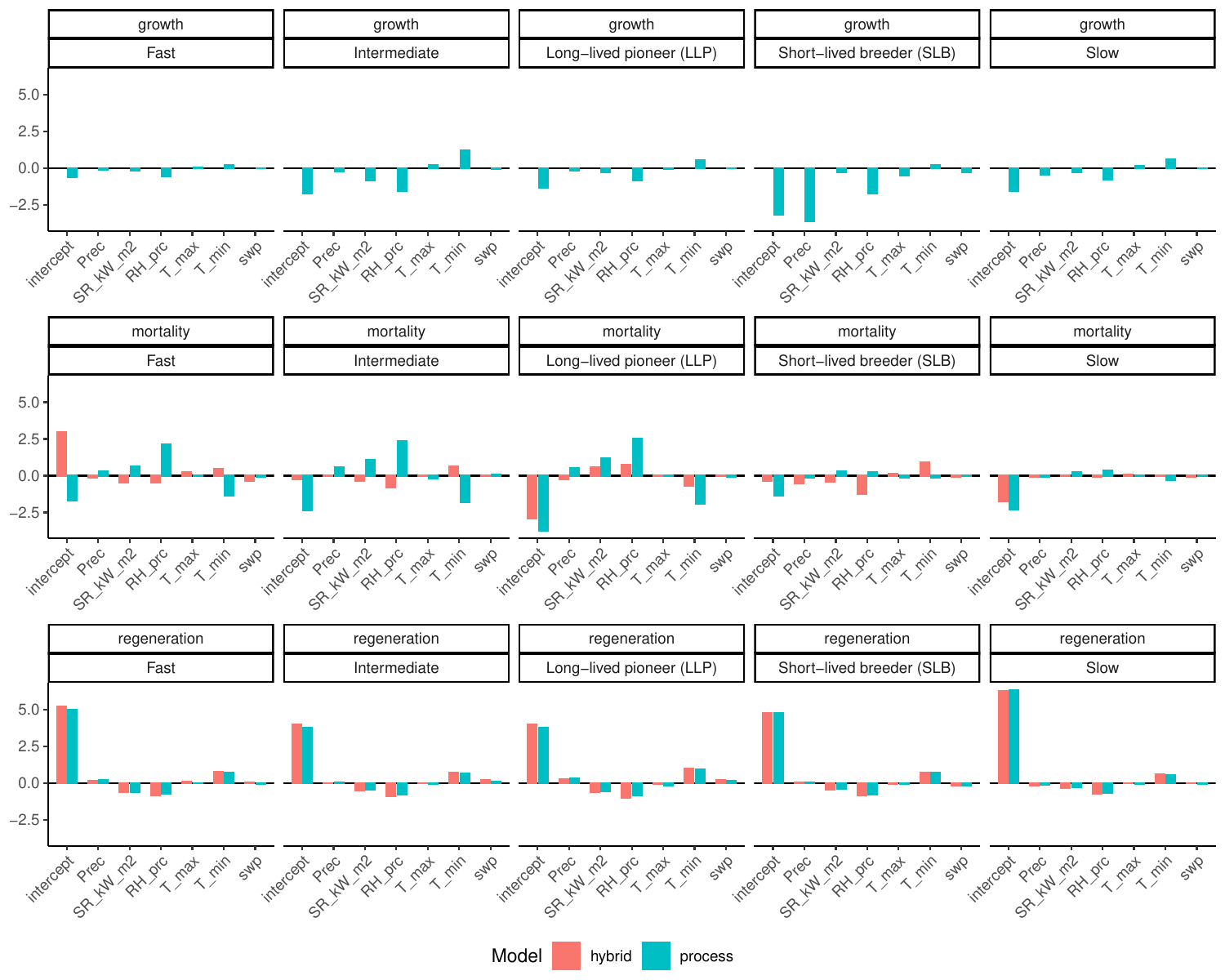}
\caption{Estimated environment parameters for Process-FINN and Hybrid-FINN calibrated on the BCI forest inventory data}
\end{figure}

\begin{figure}
\centering
\includegraphics[width=\textwidth]{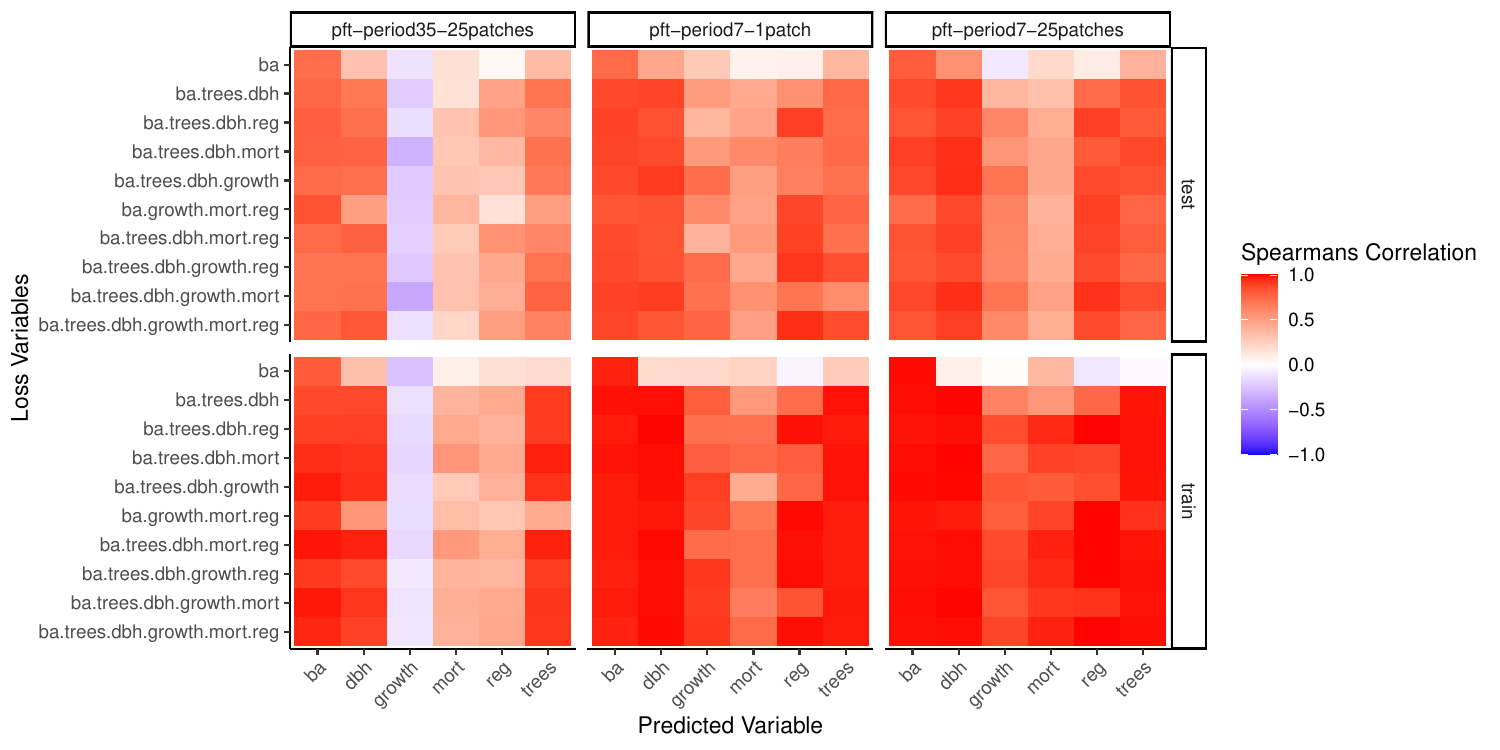}
\caption{Spearman correlations from five-fold spatial cross-validation to assess how well a model can be trained with different combinations of response variables. To ensure full coverage of each response variable, new data were simulated from the process model trained on the BCI forest inventory data. These simulations were used to calibrate models with different combinations of response variables. The correlations are averages from a spatially blocked five-fold cross-validation}
\end{figure}

\begin{figure}
\centering
\includegraphics[width=\textwidth]{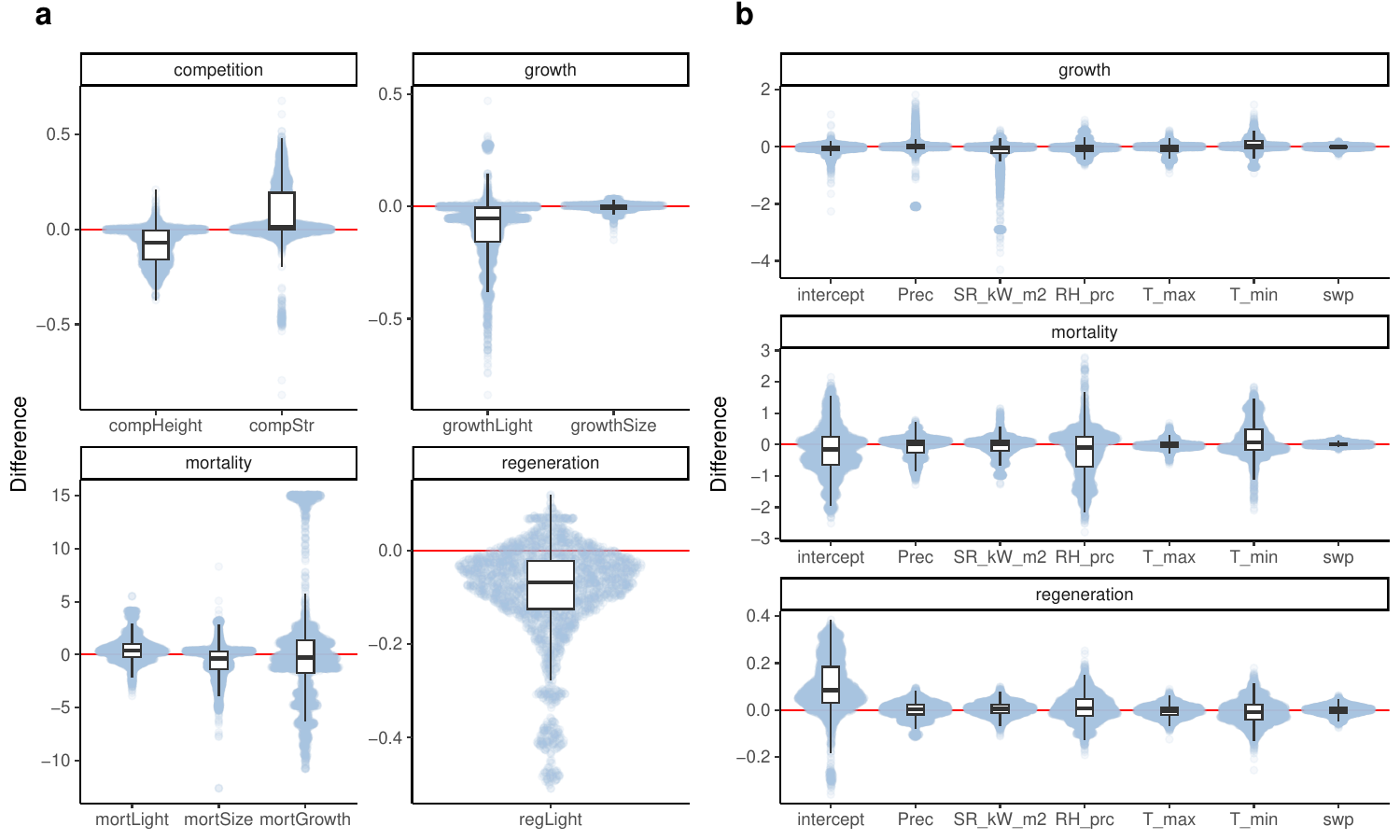}
\caption{Performance of different hybrid modeling architectures. Each hybrid model was trained on the observed BCI forest inventory data. The Spearman correlations represent the average from a spatially blocked five-fold cross-validation}
\end{figure}

\begin{figure}
\centering
\includegraphics[width=\textwidth]{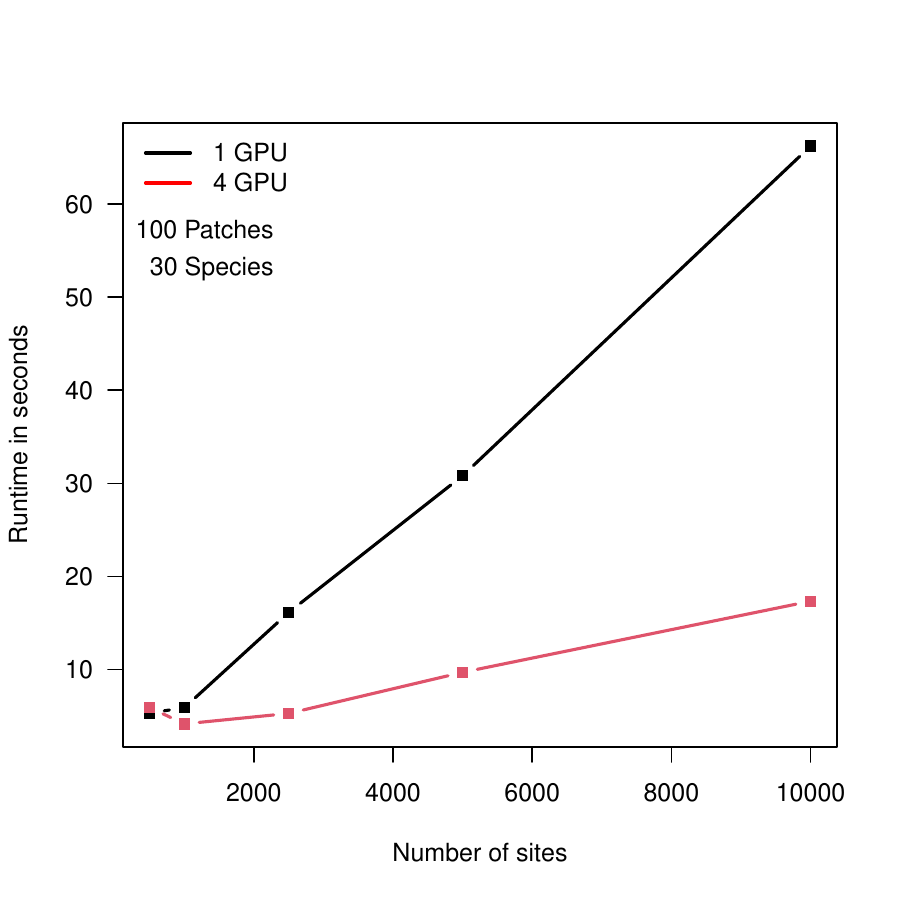}
\caption{Runtime benchmark of FINN simulations}
\end{figure}

\begin{figure}
\centering
\includegraphics[width=\textwidth]{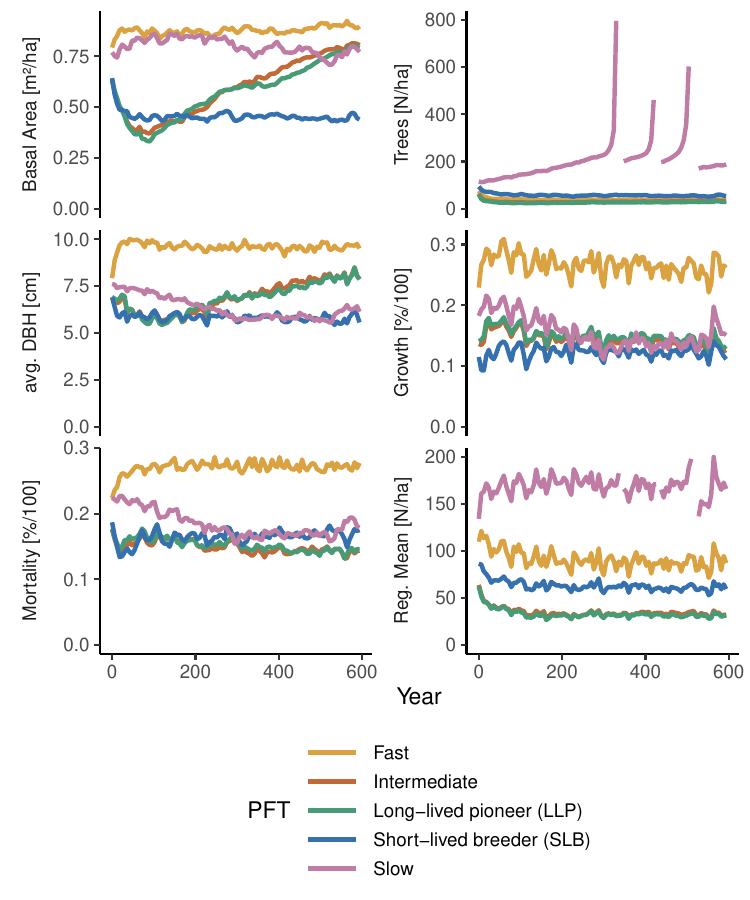}
\caption{Simulated successional trajectories of stand characteristics and demographic rates of five PFTs for  the naive DNN. Note that rates are not annual and represent changes of five year intervals. Model was trained with the BCI forest data with 7 censuses. Forest dynamics were simulated for 600 years.}
\end{figure}


\begin{table}\centering
\caption{Parameters of the Process-FINN model}
\begin{tabular}{lllllllp{5cm}} %
\toprule
Group & Process & Parameter & Unit & Type & Default & Range & Description \\
\midrule
general & all & plotsize & ha & float & 0.1 & -- & area of a patch \\
 &  & Nspecies & -- & integer & -- & 1,2,3,\ldots,n & Number of species \\
 \cmidrule(lr){2-8}
 & regeneration & dispersion & -- & float & 0.01 & -- & Dispersion parameter of the negative binomial distribution \\
\cmidrule(lr){1-8}
species & competition & compHeight & -- & float & -- & $0.3 \leq x \leq 0.7$ & height alometry defining the relation between dbh and tree height \\
 &  & compStr & -- & float & -- & $0 \leq x \leq 2$ & factor translating ba of a tree to competitive pressure for other trees \\
 \cmidrule(lr){2-8}
 & regeneration & intercept & -- & float & -- & $-5 \leq x \leq 5$ & overall effect size \\
 &  & regLight & \%/100 & float & -- & $0 < x < 1$ & light requirements of species to utilize full light (low = low light requirement; high = high light requirement) \\
 \cmidrule(lr){2-8}
 & growth & intercept & -- & float & -- & $-5 \leq x \leq 5$ & overall effect size \\
 &  & growthLight & \%/100 & float & -- & $0 < x < 1$ & light requirements of species to utilize full light (low = low light requirement; high = high light requirement) \\
 &  & growthSize & -- & float & -- & $0.01 \leq x \leq 4$ & modulating the effect of dbh on growth \\
 \cmidrule(lr){2-8}
 & mortality & intercept & -- & float & -- & $-5 \leq x \leq 5$ & overall effect size \\
 &  & mortLight & -- & float & -- & $-5 < x < 5$ & light requirements of species to utilize full light (low = low light requirement; high = high light requirement) \\
 &  & mortSize & -- & float & -- & $-5 \leq x \leq 5$ & modulating the effect of dbh on mortality \\
 &  & mortGrowth & -- & float & -- & $-5 \leq x \leq 5$ & growth dependent mortality \\
\bottomrule
\end{tabular}
\end{table}

\FloatBarrier

\bibliography{references}

\end{document}